\documentclass[a4paper,11pt]{article}
\pdfoutput=1 

\usepackage{jheppub} 
\usepackage{subcaption} 
\usepackage[T1]{fontenc} 
\usepackage{amsmath, amssymb, amsfonts, mathrsfs, slashed, bm, graphicx, xcolor, multirow}
\usepackage{color,bm}
\usepackage{adjustbox}

\usepackage{orcidlink}
\usepackage[normalem]{ulem}
\usepackage{ulem}

\usepackage[utf8]{inputenc}
\usepackage{booktabs} 
\usepackage{siunitx} 

\newcommand{\be}{\begin{equation}}
\newcommand{\ee}{\end{equation}}
\newcommand{\ba}{\begin{array}}
	\newcommand{\ea}{\end{array}}
\newcommand{\bea}{\begin{eqnarray}}
\newcommand{\eea}{\end{eqnarray}}

\newcommand{\half}{\frac{1}{2}}

\title{Probing dark-state electromagnetic form factors at future linear electron-positron colliders with polarized beams}

\author[a]{Zhong Zhang,\,\orcidlink{0009-0008-6848-9224}}
\author[a, 1]{Yu Zhang,\,\orcidlink{0000-0001-9415-8252}\note{Corresponding author.}}
\author[a, 2]{Zeren Simon Wang,\,\orcidlink{0000-0002-1483-6314}\note{Corresponding author.}}
\author[a]{and Yunlan Ji\,\orcidlink{0000-0002-9539-8837}}

\affiliation[a]{School of Physics, Hefei University of Technology, Hefei 230601,China}

\emailAdd{dayu@hfut.edu.cn}
\emailAdd{wzs@hfut.edu.cn}

\abstract{We study the electromagnetic form factors of electrically neutral dark-sector particles in this paper. Concretely speaking, we focus on operators that can lead to interactions of fermionic dark states with the Standard-Model (SM) photon, including both dimension-5 (the magnetic and electric dipole moments) and dimension-6 (the anapole moment and charge radius) operators. Correspondingly, instead of the SM-photon form factors, we employ hypercharge gauge-field form factors that can induce additional couplings to the SM $Z$-boson. Utilizing the mono-photon production channel at future linear electron-positron colliders, the International Linear Collider (ILC) and the Compact Linear Collider (CLIC), we demonstrate that tuning beam polarization allows for simultaneous enhancement of signal-event rates and suppression of dominant SM background events, i.e.~neutrino-induced processes, and hence improvement of sensitivity reach. Our analysis reveals that ILC and CLIC can probe electromagnetic form factors of dark-sector particles up to about two orders of magnitude beyond existing experimental limits. We also estimate, at the order-of-magnitude level, the validity range of the effective-field-theory (EFT) approach we adopt, finding that for the dimension-5 operators the EFT remains valid in most of the parameter regions to which the ILC and CLIC can be sensitive, while it breaks down in the entire sensitivity region for the dimension-6 operators.
}

\begin{document} 
\maketitle
\flushbottom

\section{Introduction}\label{sec:intro}

The quest to unravel the mysteries of the dark sector, a hypothetical realm beyond the Standard Model (SM) of particle physics, has been a driving force in modern physics~\cite{Antel:2023hkf}.
The dark-state particles may interact weakly with the known particles of the SM, yet may hold the key to understanding the nature of dark matter and the uncharted territories of fundamental interactions.
One of the most intriguing possibilities is that these dark states could couple to the SM photon via higher-dimensional operators~\cite{Pospelov:2000bq}.
In this work, we intend to explore observable effects at high-energy lepton colliders that could arise from such interactions, in order to broaden our understanding of the dark-sector structure.
We also emphasize that following Ref.~\cite{Chu:2018qrm} the dark-state particles studied here are not supposed to be dark-matter candidates, and thus constraints derived from relic-density observations and direct- and indirect-detection experiments are not applicable.\footnote{For existing bounds on dipole interactions of the dark matter, see e.g.~Refs.~\cite{DelNobile:2012tx,Kavanagh:2018xeh,Arina:2020mxo}.}

In this scenario, even if the dark states are entirely electromagnetically (EM) neutral, higher-dimensional effective couplings to the SM photon can still arise.
For a dark state $\chi$, modeled as a Dirac fermion, interactions with the SM photon can proceed at dimension-5 through magnetic dipole moments (MDM) and electric dipole moments (EDM), as well as at dimension-6 through anapole moment (AM) and charge radius (CR) interactions~\cite{Chu:2018qrm, Kavanagh:2018xeh}.

Since the dark state $\chi$ with EM form factors can be produced by interactions associated with the SM photon, accelerator-based experiments and astrophysical observations could serve as ideal platforms for probing these scenarios.
For instance, Ref.~\cite{Chu:2018qrm} presents the detection sensitivities at electron beam-dump experiments, including the past ones NA64~\cite{NA64:2017vtt} and mQ~\cite{Prinz:1998ua}, and the proposed LDMX~\cite{LDMX:2018cma} and BDX~\cite{BDX:2016akw} experiments.
Further, the authors of Ref.~\cite{Chu:2020ysb} investigate the possibility of probing the existence of such dark states in several representative proton beam-dump experiments, including LSND~\cite{LSND:1996jxj}, MiniBooNE-DM~\cite{MiniBooNEDM:2018cxm}, CHARM-II~\cite{CHARM-II:1989nic, CHARM-II:1994dzw}, E613~\cite{Ball:1980ojt}, as well as the projected SHiP~\cite{SHiP:2015vad} and DUNE~\cite{DUNE:2015lol, DUNE:2017pqt} experiments.
Collider-based searches for missing energy can also be exploited to probe the dark-state EM form factors.
For instance, mono-photon-signature searches at various electron-positron ($e^-e^+$) colliders have been proposed, including BaBar~\cite{Chu:2018qrm}, LEP~\cite{Fortin:2011hv, Chu:2018qrm, Zhang:2022ijx}, BESIII~\cite{Zhang:2022ijx}, Belle II~\cite{Zhang:2022ijx}, and the future experiments STCF and CEPC~\cite{Zhang:2022ijx}.
Constraints have also been derived for mono-jet searches at the LHC~\cite{Fortin:2011hv, Arina:2020mxo}.  
Furthermore, Ref.~\cite{Kling:2022ykt} studies the prospects of detecting such dark states through their scattering off the electrons in the Forward Liquid Argon Experiment detector proposed at the LHC Forward Physics Facility~\cite{Anchordoqui:2021ghd}. 
In addition, astrophysical studies, particularly those focusing on stellar-cooling constraints, provide complementary insights for the dark state $\chi$ with masses below the MeV scale~\cite{Chu:2019rok, Chang:2019xva}.

In this work, we utilize hypercharge gauge-field form factors that would induce interaction of dark states with both the SM photon and the $Z$-boson.
We focus on future high-energy electron-positron colliders with polarized beams, namely, the International Linear Collider (ILC)~\cite{ILCInternationalDevelopmentTeam:2022izu} and Compact Linear Collider (CLIC)~\cite{Robson:2018zje, Adli:2025swq}, to study the effects of tuning the incoming beams' polarization on their sensitivity reach.

This paper is structured as follows.
In section~\ref{sec:model}, we briefly describe the interactions between the dark states and photons via the higher dimensional effective couplings.
In section~\ref{sec:SignalBkg}, we present our analysis for the mono-photon signature, taking into account both signal and background events, with or without beam polarization.
The projected sensitivities to the EM form factors of the dark states at ILC and CLIC with polarized incoming beams, are presented and discussed in section~\ref{sec:sens}.
Finally, conclusions are drawn in section~\ref{sec:conclu}.

\section{Dark sectors with electromagnetic form factors}\label{sec:model}

We consider the dark state $\chi$ as a Dirac fermion which may interact effectively with the hypercharge gauge field $B_\mu$, and the effective Lagrangian is given as~\cite{Chu:2018qrm,Arina:2020mxo},
\begin{equation}
\mathcal{L}_{\chi} = \frac12\mu_{\chi}^{B}\overline{\chi}\sigma^{\mu\nu}\chi B_{\mu\nu} + \frac{i}{2}d_{\chi}^{B}\overline{\chi}\sigma^{\mu\nu}\gamma^{5}\chi B_{\mu\nu} 
- a_{\chi}^{B}\overline{\chi}\gamma^{\mu}\gamma^{5}\chi\partial^{\nu}B_{\mu\nu} + b_{\chi}^{B}\overline{\chi}\gamma^{\mu}\chi\partial^{\nu}B_{\mu\nu}.
\label{eq:lagrangian_hypercharge-field}
\end{equation}
Here, the hypercharge gauge-field strength tensor is defined as $B_{\mu\nu} \equiv \partial_\mu B_\nu - \partial_\nu B_\mu$.
The dimensional coefficients $\mu_{B}^{\chi}$ and $d_{B}^{ \chi}$ are associated with the dimension-5 MDM and EDM interactions, respectively.
They are expressed in units of the Bohr magneton $\mu_B\equiv e/(2m_e)$, where $e$ represents the electric charge and $m_e$ is the electron mass.
The symbol $\sigma_{\mu\nu}$ is defined as $\sigma_{\mu\nu} \equiv \frac{i}{2} [\gamma_\mu, \gamma_\nu]$.
The coefficients $a_{B}^{ \chi}$ and $b_{B}^{ \chi}$ correspond to the dimension-6 AM and CR interactions, respectively.
Furthermore, hypercharge form factors are linear combinations of the EM form factors and their $Z$-boson counterparts, weighted by factors related to the cosine and sine of the weak mixing angle $\theta_W$, $c_W$ and $s_W$.
Consequently, Eq~\eqref{eq:lagrangian_hypercharge-field} can be re-formulated in terms of the EM field-strength tensor $F_{\mu\nu} \equiv \partial_\mu A_\nu - \partial_\nu A_\mu$ and the $Z$-boson gauge-field strength tensor $Z_{\mu\nu} \equiv \partial_\mu Z_\nu - \partial_\nu Z_\mu$,
\begin{eqnarray}
\mathcal{L}_{\chi}&=& \frac{1}{2}\mu_{\chi}^{\gamma}\bar{\chi}\sigma^{\mu\nu}\chi F_{\mu\nu} + \frac{i}{2}d_{\chi}^{\gamma}\bar{\chi}\sigma^{\mu\nu}\gamma^{5}\chi F_{\mu\nu} - a_{\chi}^{\gamma}\bar{\chi}\gamma^{\mu}\gamma^{5}\chi\partial^{\nu}F_{\mu\nu} + b_{\chi}^{\gamma}\bar{\chi}\gamma^{\mu}\chi\partial^{\nu}F_{\mu\nu} \nonumber \\
&&+ \frac{1}{2}\mu_{\chi}^{Z}\bar{\chi}\sigma^{\mu\nu}\chi Z_{\mu\nu} + \frac{i}{2}d_{\chi}^{Z}\bar{\chi}\sigma^{\mu\nu}\gamma^{5}\chi Z_{\mu\nu} - a_{\chi}^{Z}\bar{\chi}\gamma^{\mu}\gamma^{5}\chi\partial^{\nu}Z_{\mu\nu} + b_{\chi}^{Z}\bar{\chi}\gamma^{\mu}\chi\partial^{\nu}Z_{\mu\nu},
\label{eq:lagrangian_physical-field}
\end{eqnarray}
with $C^\gamma_\chi = C^B_\chi \cos\theta_W$ and $C^Z_\chi = -C^B_\chi \sin\theta_W$, where $C_\chi=\mu_\chi, d_\chi, a_\chi,$ or $b_\chi$.

In scenarios where the energy scale is significantly below the electroweak scale, the $Z$-boson degree of freedom decouples and the effective interactions described by equation~\eqref{eq:lagrangian_hypercharge-field} or equation~\eqref{eq:lagrangian_physical-field} can thus be equivalently mapped to
\begin{equation}
\mathcal{L}_{\chi} = \frac{1}{2}\mu_{\chi}\bar{\chi}\sigma^{\mu\nu}\chi F_{\mu\nu} + \frac{i}{2}d_{\chi}\bar{\chi}\sigma^{\mu\nu}\gamma^{5}\chi F_{\mu\nu} 
- a_{\chi}\bar{\chi}\gamma^{\mu}\gamma^{5}\chi\partial^{\nu}F_{\mu\nu} + b_{\chi}\bar{\chi}\gamma^{\mu}\chi\partial^{\nu}F_{\mu\nu}.
\label{eq:lagrangian_low_energy} 
\end{equation}
The conventional EM form factors should be indicated with a $\gamma$ superscript, which will be omitted hereafter, for simplicity, unless explicitly stated otherwise.

In equation~\eqref{eq:lagrangian_hypercharge-field} we have used a simple normalization scheme for the effective operators where the relevant new-physics scales have been absorbed into single Wilson coefficients.
However, we note that the considered effective operators can arise only at loop level in weakly coupled UV-complete models~\cite{Greene:1986bm,Craig:2019wmo}; see also Ref.~\cite{Kavanagh:2018xeh} for explicit examples of UV completions.
Thus, it is non-trivial to extract the corresponding new-physics scales.
Indeed, the couplings of the dark fermions $\chi$ to the SM photon can be explicitly written as
\begin{equation}
\mu_\chi = \frac{e}{16\pi^2}\,\frac{C_{\chi\chi F}^\gamma}{\Lambda}, \quad
d_\chi = \frac{e}{16\pi^2}\,\frac{C_{\chi5\chi F}^\gamma}{\Lambda}, \quad
a_\chi = \frac{e}{16\pi^2}\,\frac{C_{\chi5\chi\partial F}^\gamma}{\Lambda^2}, \quad
b_\chi = \frac{e}{16\pi^2}\,\frac{C_{\chi\chi\partial F}^\gamma}{\Lambda^2},
\label{eq:WC_loop}
\end{equation}
where $\Lambda$ denotes the new-physics cutoff scale of the effective theory and $C_i^\gamma$'s are the new, dimensionless, corresponding Wilson coefficients with $i=\chi\chi F, \chi 5\chi F, \chi5\chi\partial F, \chi\chi\partial F$. 
At this point, it is important to point out the relevance of perturbative unitarity, which holds only for sufficiently small values of $C_i^\gamma/\Lambda$.
Concretely, for our collider analysis, we will treat values of $\mu_\chi, d_\chi, a_\chi$ and $b_\chi$ for which $C^\gamma_i\lesssim 4\pi$ with $\Lambda$ equal to the center-of-mass energy, as obeying the perturbative-unitarity requirement.
For larger values of $C_i^\gamma/\Lambda$, there may still exist experimental bounds, but perturbative calculations may not be reliable for determining these bounds.

\section{Production of the dark states at electron-positron colliders in the mono-photon channel}\label{sec:SignalBkg}

In this section, we focus on production of the dark states with the EM form factors at future $e^- e^+$ colliders.
At collider experiments, since the dark states are expected to be only feebly coupled with the SM particles, they are typically undetectable in practice, hence manifesting themselves as missing energy.
Thus, in order to detect the signal events, one has to rely on any associated final-state particles.
Thus, in this work, we will resort to the mono-photon signature for the signal-event detection.

 \subsection{Signal}\label{subsec:signal}

\begin{figure}[t]
    \centering
    \includegraphics[width=0.7\textwidth]{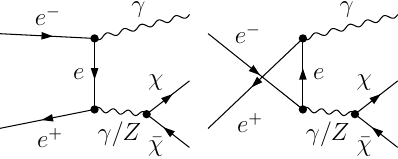}
    \caption{Feynman diagrams for the signal process $e^- e^+ \to \chi \bar{\chi} \gamma$ where the SM photon $\gamma$ is emitted from either the electron beam (left) or the positron beam (right). We omit the FSR-photon case, considering its negligible effects.}
\label{fig:signal_feynman}
\end{figure}

Via interactions with the SM photon and $Z$-boson, cf.~equation~\eqref{eq:lagrangian_physical-field}, the dark states can be pair produced in the process $e^- e^+\to \gamma/Z \to \chi \bar{\chi}$ at $e^- e^+$ colliders.
Additionally, we require an ISR (initial-state radiation) photon in the signal process in order to probe the dark states at $e^- e^+$ colliders and therefore study the following scattering process $e^- e^+ \to \chi \bar{\chi}\gamma$.\footnote{In principle, the final-state $\chi$'s can also emit a photon; however, the corresponding diagrams are at an higher order of the form factors than the considered ones, and therefore the FSR (final-state radiation)-photon processes are ignored.}
We show in figure~\ref{fig:signal_feynman} Feynman diagrams for the signal process.
The corresponding differential cross section can be approximately factorized into the cross section $\sigma_0$ of the process without photon emission, $e^- e^+\to \chi \bar{\chi}$, and the improved Altarelli-Parisi radiator function $H$~\cite{Nicrosini:1989pn,Montagna:1995wp},
\begin{equation}
    \frac{d^2\sigma}{d x_\gamma d z_\gamma}=H\left(x_\gamma, z_{\gamma}; s\right) \sigma_{0}(s_\gamma)\,,
\label{eq:diffxs}
\end{equation}
where the radiator function is given as
\begin{equation}
H\left(x_\gamma, z_{\gamma} ; s\right)=\frac{\alpha}{\pi}\frac{1}{x_\gamma} \left[\frac{1+(1-x_\gamma)^2}{1-z_{\gamma}^{2}}- \frac{x_\gamma^2}{2}\right]\,.
\label{eq:Hfunc}
\end{equation} 
Here, $s$ and $s_\gamma$ label the square of the center-of-mass (COM) energies of the $e^-e^+$ and $\chi\bar\chi$ systems, respectively.
We have $s_\gamma=(1-x_\gamma)s$, with $x_\gamma=2E_\gamma/\sqrt{s}$ being the energy fraction emitted away by ISR, where $E_\gamma$ is the energy of the ISR photon.
Further, $z_\gamma=\cos\theta_\gamma$ with $\theta_\gamma$ being the polar angle of the photon.
$\alpha$ denotes the fine structure constant.
We work in the massless limit for the incoming $e^\pm$ beams, which is a legitimate approximation at high-energy colliders, and calculate the unpolarized cross section of the $\chi$ pair production $e^- e^+\to \chi \bar\chi$ with
\bea
\label{eq:sig0}
\sigma_{0}(s)&=&\frac{\alpha}{4} \frac{1}{c_W^2}\frac{f\left(s\right)}{s^2}\sqrt{\frac{s-4 m_{\chi}^{2}}{s}}\Bigg[ c_W^2+ (g_L+g_R)\frac{s(s-M_Z^2)}{(s-M_Z^2)^2+M_Z^2\Gamma_Z^2}  \nonumber \\ 
&+&\half\frac{1}{c_W^2}(g_L^2+g_R^2)\frac{s^2}{(s-M_Z^2)^2+M_Z^2\Gamma_Z^2}\Bigg],
\eea
where $g_L=-\half+\sin\theta_W^2$ and $g_R=\sin\theta_W^2$, $M_Z$ and $\Gamma_Z$ are the mass and decay width of the $Z$-boson, respectively, and $m_\chi$ labels the mass of $\chi$.

In equation~\eqref{eq:sig0}, the function $f(s)$ is unique for each high-dimension operator we study and can be expressed respectively as
\begin{align}
\mathrm{MDM}: f\left(s\right) &= \frac{2}{3} \mu_{\chi}^2 s^{2}\left(1+\frac{8 m_{\chi}^{2}}{s}\right), \label{eq:MDM} \\
\mathrm{EDM}: f\left(s\right) &= \frac{2}{3} d_{\chi}^2 s^{2}\left(1-\frac{4 m_{\chi}^{2}}{s}\right), \label{eq:EDM} \\
\mathrm{AM}: f\left(s\right) &= \frac{4}{3} a_{\chi}^2 s^{3}\left(1-\frac{4 m_{\chi}^{2}}{s}\right), \label{eq:AM} \\
\mathrm{CR}: f\left(s\right) &= \frac{4}{3} b_{\chi}^2 s^{3}\left(1+\frac{2 m_{\chi}^{2}}{s}\right). \label{eq:CR}
\end{align}

If  the incoming $e^\mp$ beams are polarized, the  polarized cross sections $\sigma_0^{\lambda_{e^{-}} \lambda_{e^{+}}}$ for the $e^- e^+ \to \chi \bar{\chi}$ process can be expressed as,
\bea
\sigma_{0}^{-+}(s)&=&\frac{\alpha}{4} \frac{f\left(s\right)}{s^2}\sqrt{\frac{s-4 m_{\chi}^{2}}{s}}\Bigg[2 c_W^2+ 4 g_{L}\frac{s(s-M_Z^2)}{(s-M_Z^2)^2+M_Z^2\Gamma_Z^2}  \nonumber \\ 
&+&\frac{2}{c_W^2}g_{L}^2\frac{s^2}{(s-M_Z^2)^2+M_Z^2\Gamma_Z^2}\Bigg],
\eea
\bea
\sigma_{0}^{+-}(s) = \sigma_{0}^{-+}(s) \text{ with $g_L$ replaced by $g_R$,}
\eea
and 
\bea
\sigma_{0}^{++}(s)=\sigma_{0}^{--}(s)=0,
\eea
where $\lambda_{e^-/e^+}$ is the electron/positron helicity and can be either $+1$ or $-1$.
To understand the notation better, taking as an example, $\sigma_0^{+-}$ stands for $\lambda_{e^-}=+1$ corresponding to a fully right-handed polarized electron beam, and $\lambda_{e^+}=-1$ corresponding to a fully left-handed polarized positron beam.

Since the ISR photon has no impact on the polarization, the fully polarized cross section $\sigma^{ \lambda_{e^{-}}\lambda_{e^{+}} }$ for the $e^- e^+\to \chi \bar\chi \gamma$ process can be directly obtained from  $\sigma_0^{\lambda_{e^{-}}  \lambda_{e^{+}} }$ combined with equation~\eqref{eq:diffxs} and equation~\eqref{eq:Hfunc}.

The cross sections for electron- and positron-beam polarization rates $P_{e^-}$ and $P_{e^+}$, respectively, are given (in helicity basis) by,
\bea
\sigma\big(P_{e^-}, P_{e^+} \big)
=
  \sum_{  \lambda_{e^{-}}, \lambda_{e^{+}}= \pm 1}
\frac {1 + \lambda_{e^{-}} P_{e^-}} 2
  \frac {1 + \lambda_{e^{+}} P_{e^+}} 2
  \sigma^{\lambda_{e^{-}}\lambda_{e^{+}}},
\label{eq:sigmaP}
\eea
where $P_{e^-}$ and $P_{e^+}$ are defined as $P_{e^\mp} \equiv (\phi_{e^\mp_\uparrow} - \phi_{e^\mp_\downarrow}) / (\phi_{e^\mp_\uparrow} + \phi_{e^\mp_\downarrow})$ with $\phi_{e^\mp_{\uparrow,\downarrow}}$ being the electron/positron fluxes with positive ($\uparrow$) and negative ($\downarrow$) helicities, respectively.
Then the prefactor $(1 + \lambda_{e^{\mp}}P_{e^\mp}) / 2$ is the weight for each beam-polarization component.

In order to avoid collinear and infrared singularities (which are also present in the Altarelli-Parisi radiator function, cf.~equation~\eqref{eq:Hfunc}), we impose the following kinematical cuts on the final-state photon~\cite{Habermehl:2020njb},
\begin{equation}
p_T^\gamma>1\:\mathrm{GeV}, 4^\circ<\theta_\gamma<176^\circ \:,  \label{eq:initial-cut}  
\end{equation}
where $p_T^\gamma$ and $\theta_\gamma$ are respectively the transverse momentum and polar angle of the photon.
By integrating the differential cross section in equation~\eqref{eq:diffxs} with the cuts listed in equation~\eqref{eq:initial-cut}, we derive the total cross sections at the $e^-e^+$ colliders, for the process $e^-e^+\to\chi\bar{\chi}\gamma$ induced by the EM form factors of the dimension-5 MDM and EDM operators or the dimension-6 AM and CR operators.

\begin{figure}[t]
    \centering
    \includegraphics[width=0.495\textwidth]{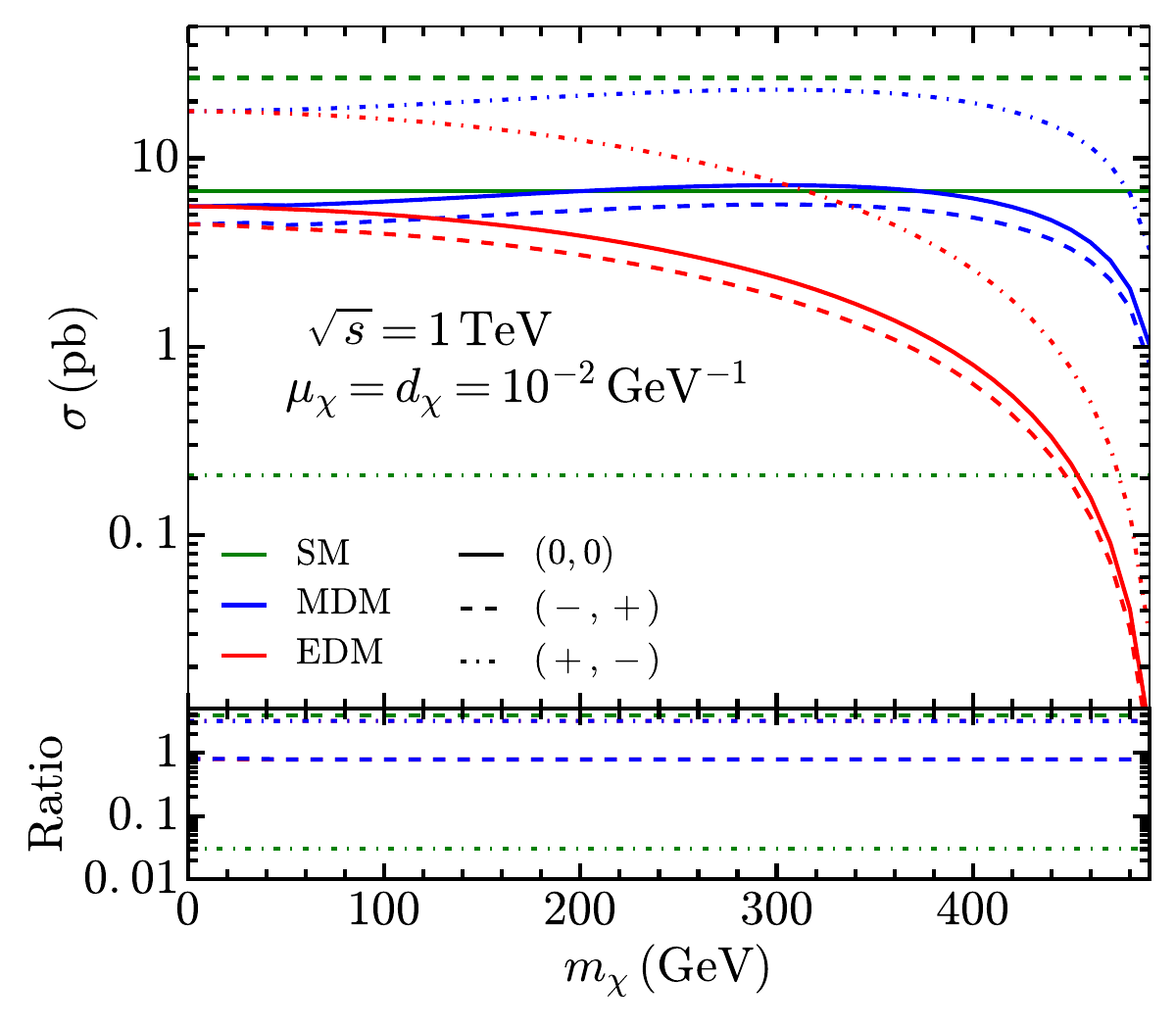}
    \includegraphics[width=0.495\textwidth]{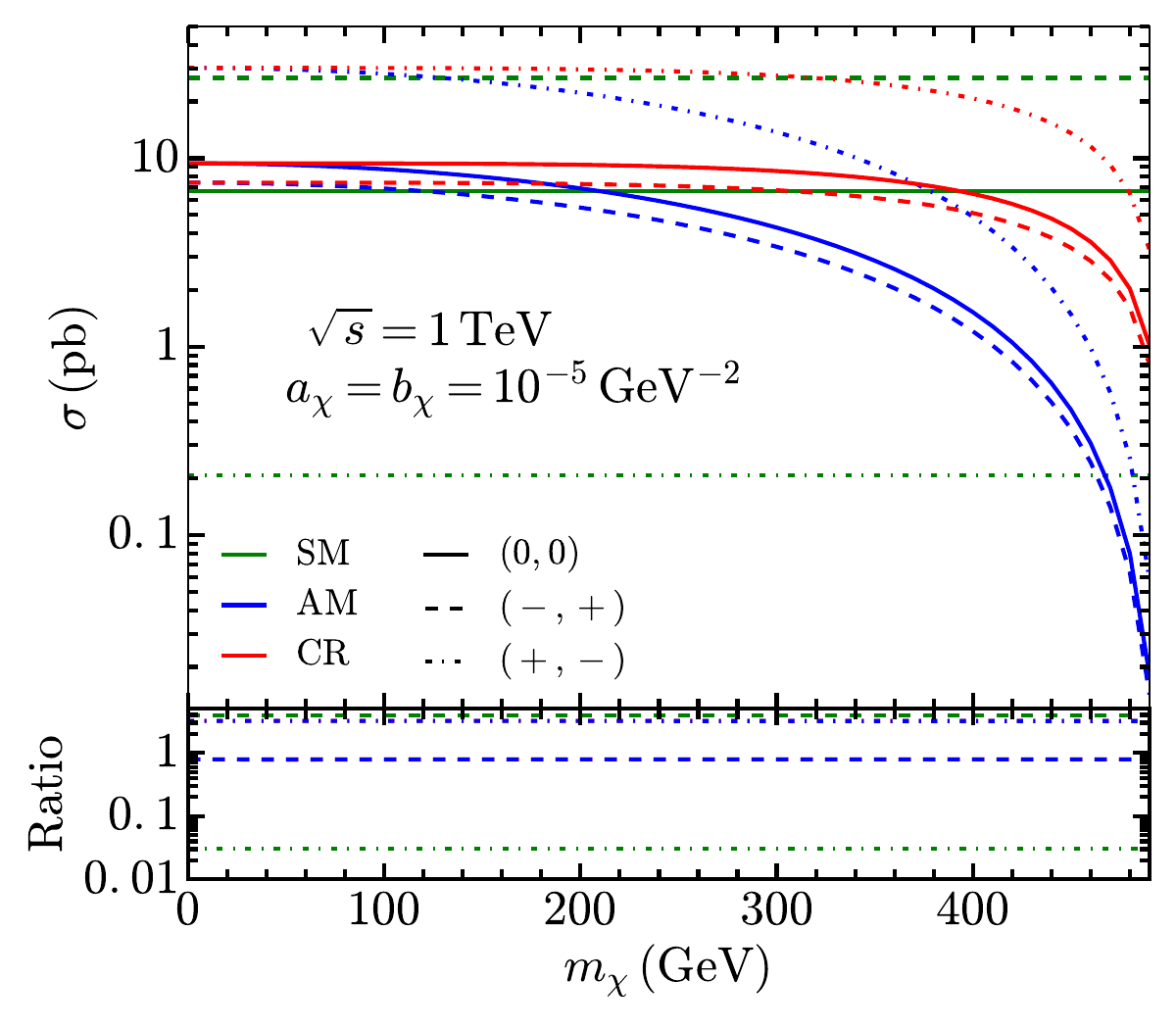}
    \caption{Mono-photon signal cross sections as functions of $m_\chi$ at an $e^-e^+$ collider with $\sqrt{s} = 1$ TeV.
    The benchmark parameters are $d_\chi = \mu_\chi = 10^{-2}$ GeV$^{-1}$ and $a_\chi = b_\chi = 10^{-5}$ GeV$^{-2}$.
    The left (right) panel shows results for MDM and EDM (AM and CR).
    Line styles indicate different polarization configurations: solid $(0, 0)$ for ($P_{e^-} = 0\%$, $P_{e^+} = 0\%$), dashed $(-, +)$ for ($P_{e^-} = -100\%$, $P_{e^+} = +100\%$), and dot-dashed $(+, -)$ for ($P_{e^-} = +100\%$, $P_{e^+} = -100\%$).
    ``SM'' labels the curves for the irreducible SM background process $e^- e^+\to \nu \bar{\nu}\gamma$; see section~\ref{subsec:background} for detail.
    In addition, we show in the lower parts of both panels the ratios of cross sections with the polarized beam setups to those with the unpolarized one; we note that the ratios for the MDM and EDM operators in the left panel (the AM and CR operators in the right panel) overlap with each other.
    }
\label{fig:sig-mchi-1tev}
\end{figure}

In figure~\ref{fig:sig-mchi-1tev}, we present the dependence of the signal cross sections on $m_\chi$ at an $e^-e^+$ collider with a COM energy of 1 TeV.
We consider three extreme cases of polarization configurations: ``$(0, 0)$'' for $(P_{e^-} = 0\%, P_{e^+} = 0\%)$, ``$(+, -)$'' for $(P_{e^-} = +100\%, P_{e^+} = -100\%)$, and ``$(-, +)$'' for $(P_{e^-} = -100\%, P_{e^+} = +100\%)$.
We observe that for $m_\chi^2/s \to 0$ and $\mu_\chi=d_\chi$, the cross sections of the signal process mediated via the dimension-5 MDM and EDM operators are almost identical.
This behavior can be easily understood by comparing equation~\eqref{eq:MDM} and equation~\eqref{eq:EDM}.
For the dimension-6 AM and CR operators, we observe the same behavior, cf.~equation~\eqref{eq:AM} and equation~\eqref{eq:CR}.
Moreover, as shown in figure~\ref{fig:sig-mchi-1tev}, the signal cross section rapidly decreases with increasing $m_\chi$, except for the MDM case where the cross section gets mildly enhanced first before declining.
This behavior can be understood by differentiating equation~\eqref{eq:sig0} with respect to $m_\chi$.

\begin{figure}[t]
\centering
\includegraphics[width=0.495\textwidth]{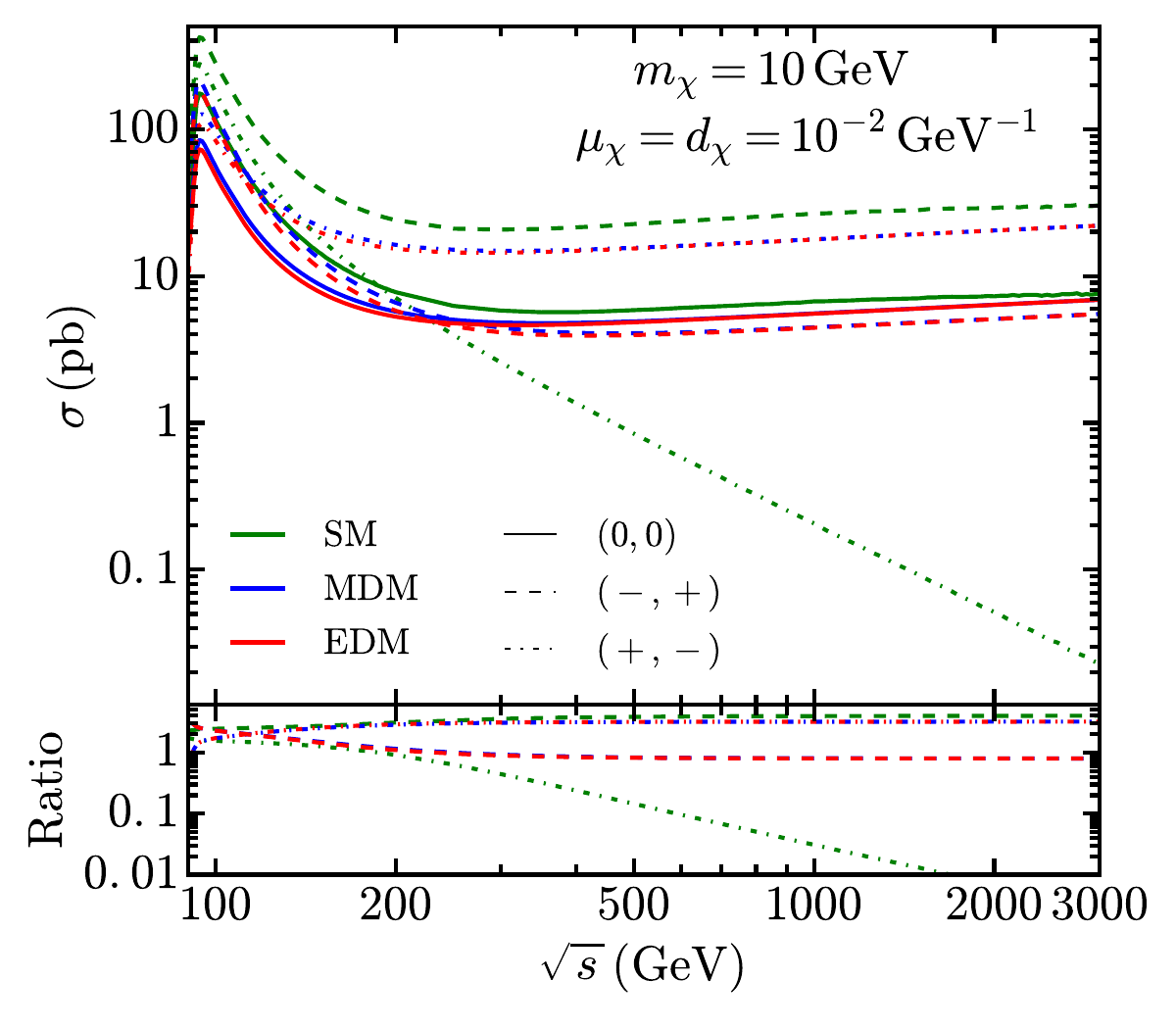}
\includegraphics[width=0.495\textwidth]{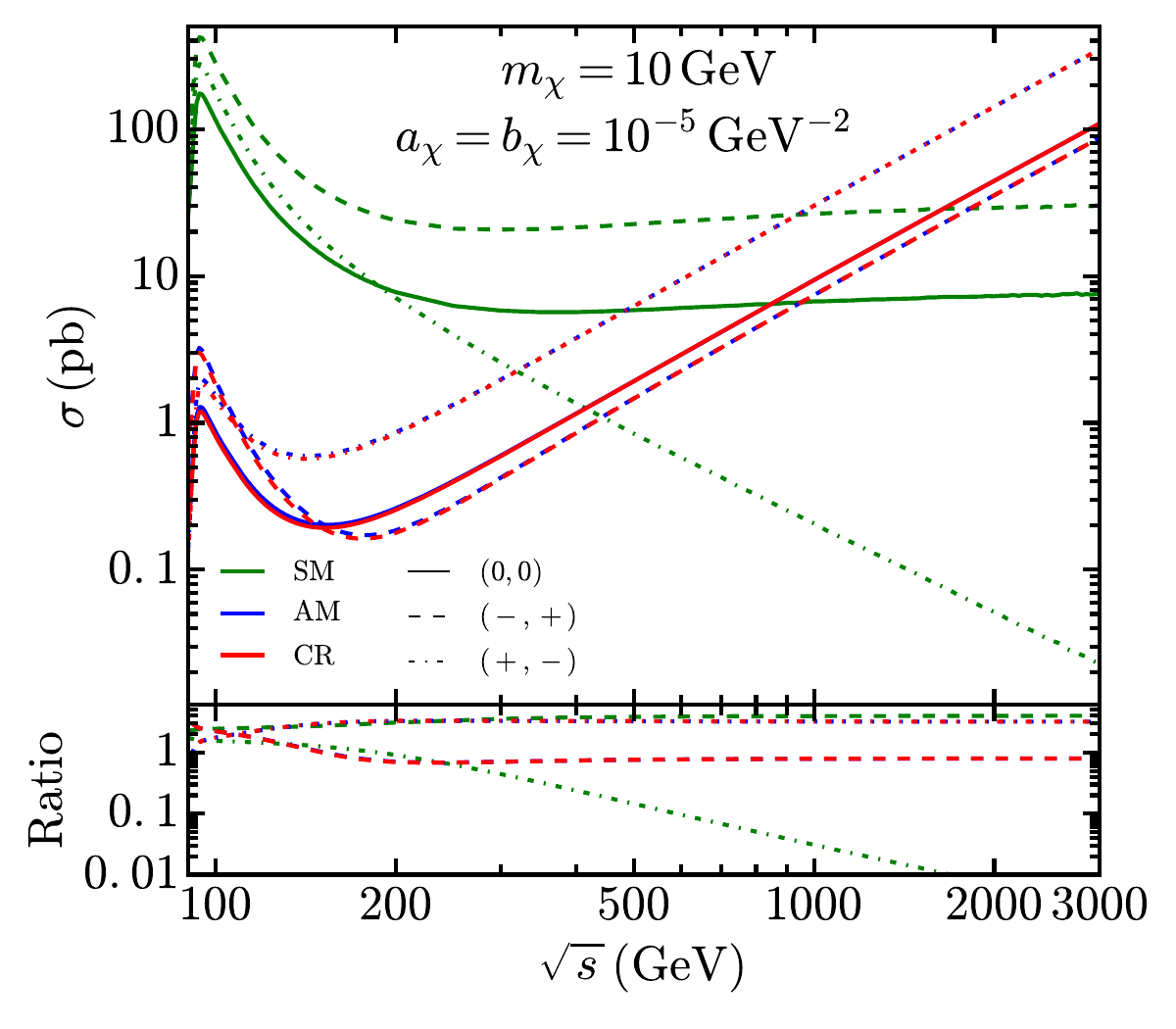}
\caption{Mono-photon signal cross sections as functions of $\sqrt{s}$ at an $e^-e^+$ collider, with $m_\chi$ fixed at 10 GeV.
The format follows figure~\ref{fig:sig-mchi-1tev}.
}
\label{fig:sig-s-m}
\end{figure}

In figure~\ref{fig:sig-s-m}, we plot the signal cross sections as functions of the COM energy $\sqrt{s}$ at an $e^-e^+$ collider, for $m_\chi = 10$ GeV.
The considered polarization configurations are the same as those in figure~\ref{fig:sig-mchi-1tev}.
Figure~\ref{fig:sig-s-m} indicates that for $\sqrt{s}\gtrsim 150$ GeV, the signal cross sections with the dimension-5 MDM and EDM operators do not depend strongly on $\sqrt{s}$, while those with the dimension-6 operators grow rapidly with $\sqrt{s}$.
This can be understood by the different scaling behaviors of the cross sections with respect to $s$; see equation~\eqref{eq:MDM} -- equation~\eqref{eq:CR}.
In addition, the peak observed at $\sqrt{s} \approx 94$ GeV in the figure is due to the $Z$-boson resonance effect.

Figure~\ref{fig:sig-mchi-1tev} and figure~\ref{fig:sig-s-m} show that in most cases, the dimension-5 operators EDM and MDM exhibit similar characteristics, as do the dimension-6 operators AM and CR.
Therefore, in order to simplify the analysis and highlight the main conclusions, we will primarily present the results for EDM and AM in the subsequent analysis.

\subsection{Background}\label{subsec:background}

For the mono-photon signature at electron-positron colliders, the background sources can be either reducible or irreducible.

\begin{figure}[t]
\centering
\includegraphics[width=0.75\textwidth]{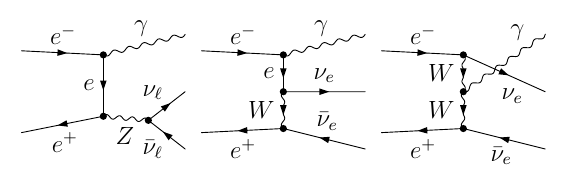}\\
\includegraphics[width=0.5\textwidth]{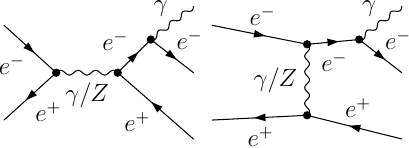}
\caption{Upper: The Feynman diagrams for the  process $e^- e^+ \to \nu_\ell\bar{\nu}_\ell\gamma$, where $\nu_\ell=\nu_e,\nu_\mu,\nu_\tau$ are the SM neutrinos. 
Lower: The Feynman diagrams for the  radiative Bhabha scattering $e^-e^+\to e^-e^+\gamma$.
Diagrams with photon radiated by the positron are included in the physics analysis but not shown here.}
\label{fig:bkg-feyndia}
\end{figure}

The irreducible SM background originates from the process $e^- e^+ \to \nu_\ell\bar{\nu}_\ell\gamma$, where $\nu_\ell=\nu_e,\nu_\mu,\nu_\tau$ labels the SM neutrinos.
The corresponding Feynman diagrams are shown in the upper panel of figure~\ref{fig:bkg-feyndia}.
For the electron neutrino, all the listed diagrams, i.e.~both $Z$-boson and $W$-boson diagrams, contribute; while for the muon and tau neutrinos, only the first diagram, i.e.~only $Z$-boson diagrams, contribute.

The reducible background arises from one visible photon in the final state associated with other SM particles that are undetected because of detector limitations.
The major contribution of reducible background stems from the radiative Bhabha scattering process
$e^-e^+\to e^-e^+\gamma$, in which the final-state electron and positron both travel in the very forward region and thus escape detection.
Since there are several singularity configurations of kinematics in the cross section of this process~\cite{Berends:1983fs, Tobimatsu:1985vd}, it may give a sizable contribution as reducible background.
This reducible background can be suppressed to the per mille level, by applying vetoes on the objects reconstructed in the electromagnetic calorimeters located in the very forward region (BeamCal), achieving a suppression factor of 0.23\%~\cite{Habermehl:2020njb, Habermehl:2018yul}. 
As shown in Ref.~\cite{Habermehl:2020njb}, the reducible background is more than 2 orders of magnitude smaller than the irreducible one and therefore can be safely neglected.
Consequently, unless otherwise specified, our discussion focuses on the irreducible background.
In figure~\ref{fig:sig-s-m}, we also present the dependence of the cross section of the irreducible background process $e^-e^+\to \nu \bar\nu \gamma$ on $\sqrt{s}$ at an $e^-e^+$ collider with the three identical benchmark polarization configurations as the signal process.
One observes that there is a peak observed around 94 GeV, which arises from the $Z$-boson resonance effect.

\subsection{Effect of polarization}\label{subsec:polarization}

Previously, we have shown in figure~\ref{fig:sig-mchi-1tev} and figure~\ref{fig:sig-s-m} the signal and background cross sections in three extreme cases of polarization configurations.
The ratios of cross sections with the fully polarized  configurations $(+,-)$ and $(-,+)$  to those with the unpolarized one  $(0,0)$ are also shown in the lower parts.
We find that different beam-polarization configurations can affect significantly the scattering cross sections.
Compared to the unpolarized case, at an 1-TeV $e^-e^+$ collider with fully right-handed polarized electron beam and fully left-handed polarized positron beam (the $(+,-)$ beam-polarization configuration), the signal cross sections with the dimension-5 (dimension-6) operators can be enhanced by a factor of more than 3, while the irreducible background can be lowered by a factor of $\sim 32$.  
Figure~\ref{fig:sig-s-m} shows that with the increment of $\sqrt{s}$ from 91 GeV to 3 TeV, the ratio of the signal cross sections with the $(+,-)$ beam-polarization configuration to the unpolarized case increases approximately from 0.85 to 3.2. On the other hand, for the irreducible background, the cross section decreases rapidly with increasing $\sqrt{s}$ in the $(+,-)$ case for $\sqrt{s}\gtrsim 200$ GeV.
Concretely speaking, the ratio of the cross section in $(+,-)$ case to the one in $(0,0)$ case can decrease to about $0.3\%$ at $\sqrt{s}\simeq 3$ TeV.
This is because that only the left-handed electron and right-handed positron can participate in the SM charged-current interactions (the $W$-boson-mediated diagrams), and that in this case the larger the collision energy, the more dominant the contribution of the $W$-boson-mediated diagrams become.
These findings imply that it is possible to enhance the signal-to-background ratio and hence sensitivity reach, by tuning the beam polarizations.

\begin{table}[t]
    \centering
    \begin{tabular}{ccccc}
        \toprule
        Collider &  $\sqrt{s}$ (GeV) & $P_{e^{-}/e^{+}}[\%]$  & $(--, -+, +-, ++) (\%)$ & $\mathcal{L}_{\text{int}}$ (ab$^{-1}$) \\
        \midrule
        ILC  &  91  & $\pm 80/\pm 30$ & (10, 40, 40, 10) & 0.1 \\
             &  250 & $\pm 80/\pm 30$ & (5, 45, 45, 5)   & 2   \\
             &  350 & $\pm 80/\pm 30$ & (5, 68, 22, 5)   & 0.2 \\
             &  500 & $\pm 80/\pm 30$ & (10, 40, 40, 10) & 4   \\
             &  1000 & $\pm 80/\pm 20$ & (10, 40, 40, 10) & 8   \\
        \midrule
        Collider &  $\sqrt{s}$ (GeV) & $P_{e^{-}}[\%]$ & $(-, +) (\%)$ & $\mathcal{L}_{\text{int}}$ (ab$^{-1}$) \\
        \midrule
        CLIC &  380  & $\pm 80$      & (50, 50) & 4.3   \\
             &  1500 & $\pm 80$      & (80, 20) & 4 \\
             &  3000 & $\pm 80$      & (80, 20) & 5   \\
        \bottomrule
    \end{tabular}
    \caption{Summary of the ILC and CLIC electron-positron collision modes, listing the COM energy $\sqrt{s}$, polarization mode, operation-time proportion of each polarization mode, and total integrated luminosity $\mathcal{L}_{\text{int}}$ for each $\sqrt{s}$.}
    \label{tab:pol-coll}
\end{table}

\begin{table*}[!t]
    \centering
    \resizebox{\textwidth}{!}{
    \begin{tabular}{lccccccc} 
        \toprule
        {Process Type} & {Unpolarized Cross Section (fb)} & {Polarization} & \multicolumn{4}{c}{Polarized Cross Section (fb)} \\
        & & \((P_{e^-}, P_{e^+})\) & \((+,+)\) & \((+,-)\) & \((-,+)\) & \((-,-)\) \\
        \midrule
        \(\nu\overline{\nu}\gamma\)   &              \num{6.62E3}& {(80, 20)} & \num{1.65E3} & \boldsymbol{$1.16\times 10^3$} & \num{1.43E4} & \num{9.55E3} \\ 
        \midrule
        EDM                          & 55.3         & {(80, 20)} & 66.3         & \textbf{97.3}        & 31.0         & 25.6         \\
        \midrule
        AM                           & 93.6         & {(80, 20)} & 112.6         & \textbf{165.2}       & 51.9         & 44.7        \\
        \bottomrule
    \end{tabular}}
    \caption{Comparison of the cross sections of the irreducible SM background and signal processes, with unpolarized beams and various polarization-sign choices for $|P(e^-, e^+)| = (80\%, 20\%)$ at ILC with $\sqrt{s}=1$ TeV.
    The cut conditions given in equation~\eqref{eq:initial-cut} are imposed.
    For the signal, benchmark values of $m_\chi = 10\,\text{GeV}$, $d_\chi = \mu_\chi = 10^{-3}$ GeV$^{-1}$, and $a_\chi = b_\chi = 10^{-6}$ GeV$^{-2}$ are chosen.
    The numbers in bold highlight the optimal helicity-orientation choice for each operator type.}
    \label{table:difpol-SM-sig-ILC}
\end{table*}

\begin{table*}[!t]
    \centering
    \resizebox{\textwidth}{!}{
    \begin{tabular}{lccccc} 
        \toprule
        {Process Type} & {Unpolarized Cross Section (fb)} & {Polarization} & \multicolumn{2}{c}{Polarized Cross Section (fb)} \\
        & & \((P_{e^-}, P_{e^+})\) & \((+,0)\) & \((-,0)\) \\
        \midrule
        \(\nu\overline{\nu}\gamma\)   &   \num{7.64e3} 
          & {(80, 0)} & \boldsymbol{$1.52\times 10^3$} & \num{13.6e3}  \\ 
        \midrule
        EDM                          & \num{68.8}        & {(80, 0)} & \boldsymbol{$101.9$}        & \num{35.8}         \\
        \midrule
        AM                           & \num{1.09e3}         & {(80, 0)} & \boldsymbol{$1.62\times 10^3$}               & \num{567.6}       \\
        \bottomrule
    \end{tabular}}
    \caption{Comparison of the cross sections of the irreducible SM background and signal processes, with unpolarized beams and various polarization-sign choices for $|P(e^-, e^+)| = (80\%, 0\%)$ at CLIC with $\sqrt{s}=3$ TeV.
    The cut conditions given in equation~\eqref{eq:initial-cut} are imposed.
    For the signal, benchmark values of $m_\chi = 10\,\text{GeV}$, $d_\chi = \mu_\chi = 10^{-3}$ GeV$^{-1}$, and $a_\chi = b_\chi = 10^{-6}$ GeV$^{-2}$ are chosen.
    The numbers in bold highlight the optimal helicity-orientation choice for each operator type.
    }
    \label{table:difpol-SM-sig-CLIC}
\end{table*}

We summarize the planned operation modes at ILC~\cite{ILCInternationalDevelopmentTeam:2022izu} and CLIC~\cite{Robson:2018zje, Adli:2025swq} with polarized beams in table~\ref{tab:pol-coll}, listing the COM energy $\sqrt{s}$, polarization mode, operation-time proportion of each polarization mode, and total integrated luminosity $\mathcal{L}_{\text{int}}$ for each $\sqrt{s}$.
With COM energies up to 500 GeV at ILC, the electron beam is supposed to be $\pm 80 \%$ polarized while the positron is planned for $\pm 30\%$ polarization.
With $\sqrt{s}=1$ TeV, the positron-beam polarization at ILC is set to be $\pm 20\%$.
At CLIC, with $\sqrt{s}= 380, 1500,$ and 3000 GeV, $P_{e^-}=\pm 80\%$ while the positron beam should be unpolarized.

In table~\ref{table:difpol-SM-sig-ILC}, we list the cross sections of both background and signal processes at ILC with $\sqrt{s}=1$ TeV, for various polarization-sign combinations of $|P(e^-, e^+)| = (80\%, 20\%)$, where we have applied the cut conditions given in equation~\eqref{eq:initial-cut}.
Here, we have used the Monte-Carlo (MC) simulation tool \texttt{MadGraph5}~\cite{Alwall:2014hca} to generate the SM irreducible background events and calculate the corresponding cross sections.
As for the signal process, we have computed the cross sections by integrating equation~\eqref{eq:diffxs}.

As shown in table~\ref{table:difpol-SM-sig-ILC}, tuning the beam polarizations has a significant impact on both the signal and background cross sections.

The signal and background cross sections vary to different degrees if we tune the beam polarizations.
Specifically, for both EDM and AM signals, the $(P_{e^-}, P_{e^+}) = (+80\%, -20\%)$ helicity-orientation configuration, among all the considered ones, is predicted to have both the strongest enhancement on the signal process and suppression on the background process.

We have also performed a similar analysis for the signal and background cross sections at CLIC with $\sqrt{s}=3$ TeV; the results are listed in table~\ref{table:difpol-SM-sig-CLIC}.
We conclude that for this collider setup, the polarization scheme $(P_{e^-}, P_{e^+}) = (+80\%, 0\%)$ is predicted to have signal (background) cross sections larger (smaller) than the scheme $(P_{e^-}, P_{e^+}) = (-80\%, 0\%)$, simultaneously.

\begin{figure}[t]
    \centering
    \includegraphics[width=0.495\textwidth]{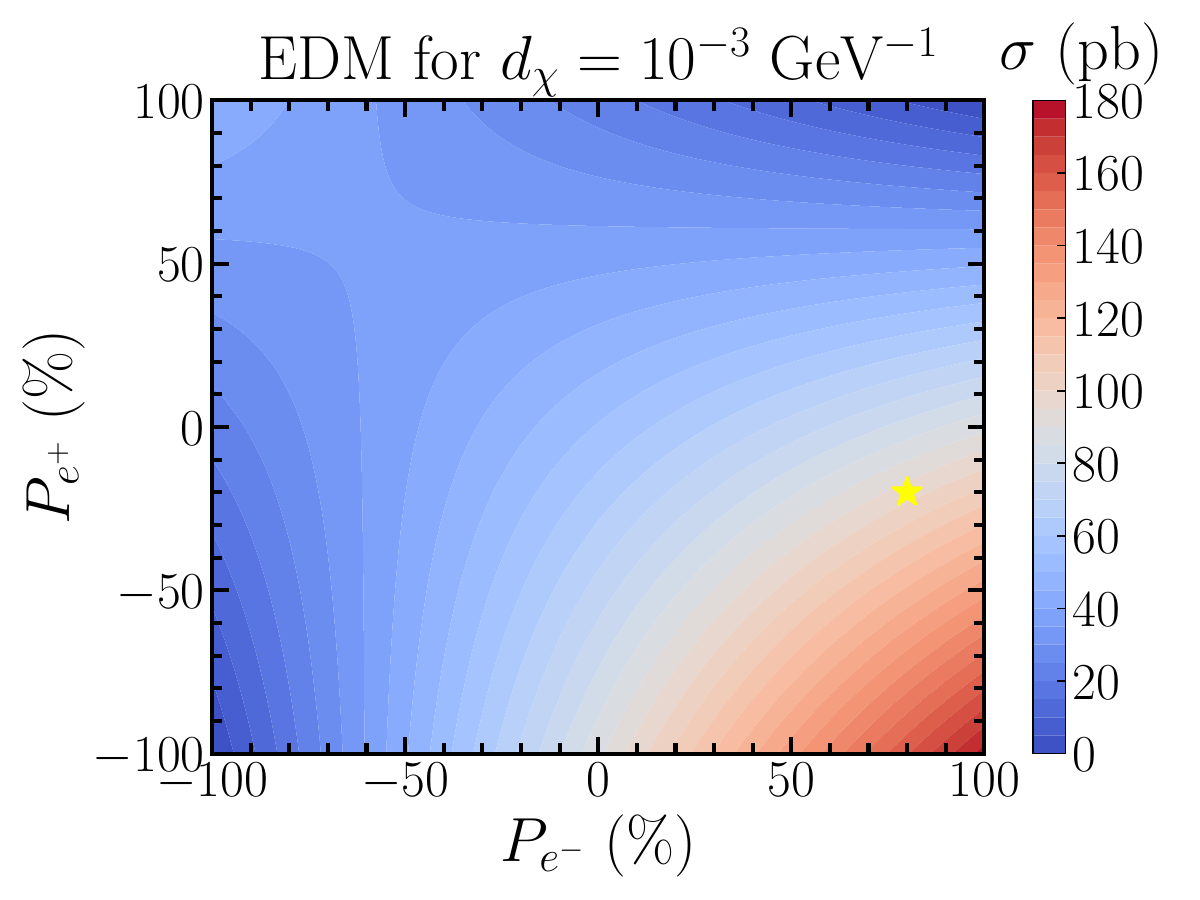}
    \includegraphics[width=0.495\textwidth]{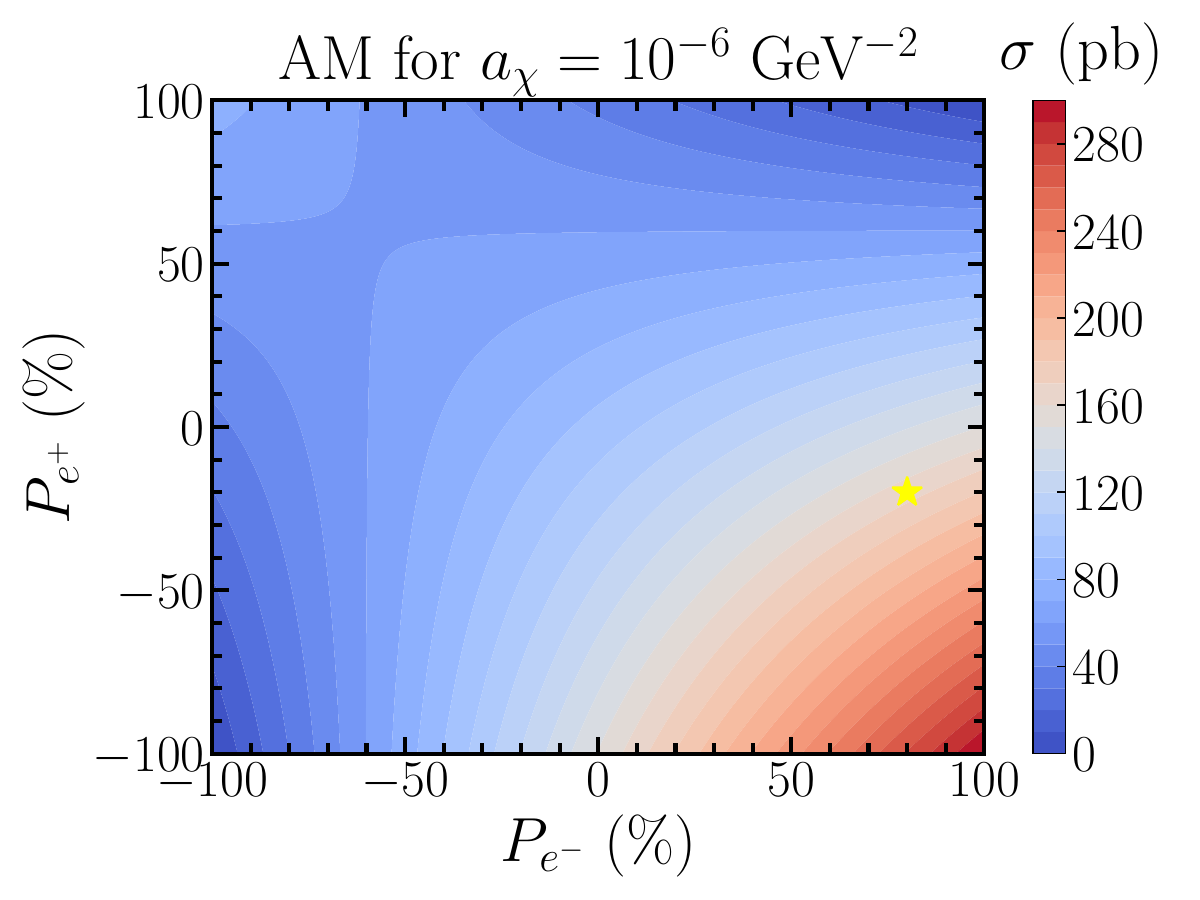}
    \includegraphics[width=0.495\textwidth]{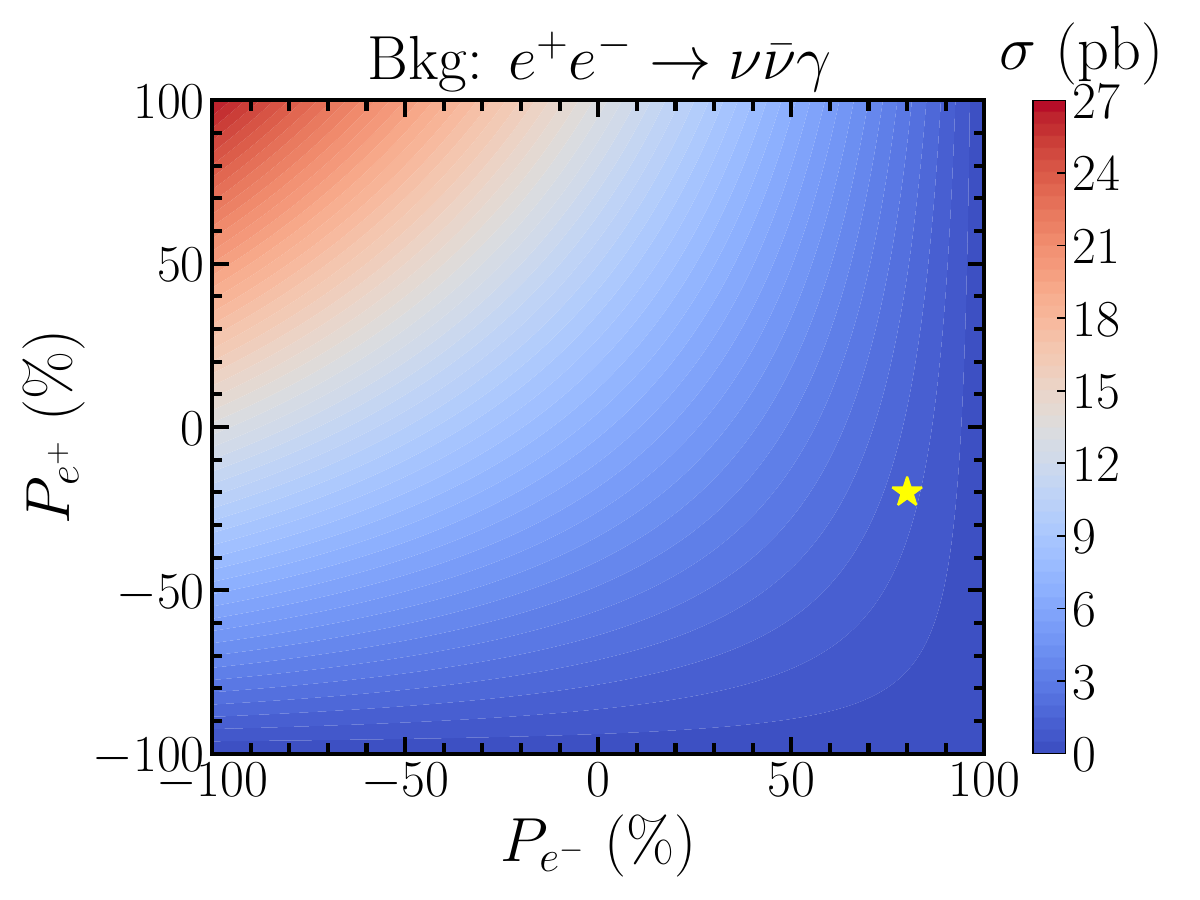}
    \caption{The scattering cross sections of the signal process with the EDM (upper left) and AM (upper right), and the neutrino-background process (lower), as functions of the incoming beams' polarizations $P_{e^\pm}$, with benchmark parameters $\sqrt{s} = 1\,\text{TeV}, m_\chi = 10\,\text{GeV}, d_\chi = 10^{-3}$ GeV$^{-1}$, and $a_\chi = 10^{-6}$ GeV$^{-2}$.
    The kinematic requirements on the photon as given in equation~\eqref{eq:initial-cut} are included.
    The yellow star in each plot marks the $(P_{e^-},P_{e^+})=(+80\%, -20\%)$ position corresponding to the polarization choice for each operator type that is predicted to have the largest signal-to-background cross-section ratio among all four possible helicity-orientations, cf.~table~\ref{table:difpol-SM-sig-ILC}.}
    \label{fig:EDM_AM_SM_polarization}
\end{figure}

Before proceeding to perform collider analyses and present sensitivity results, we show in figure~\ref{fig:EDM_AM_SM_polarization} the signal- and background-processes' cross sections in the $P_{e^+}$ vs.~$P_{e^-}$ plane, with the cuts listed in equation~\eqref{eq:initial-cut} on the final photon included.
In these plots, we have fixed $\sqrt{s}=1$ TeV, $m_\chi=10$ GeV, $d_\chi=10^{-3}$ GeV$^{-1}$, and $a_\chi=10^{-6}$ GeV$^{-2}$.
Figure~\ref{fig:EDM_AM_SM_polarization} helps to understand better, compared to our chosen $(P_{e^-}, P_{e^+}) = (+80\%, -20\%)$ configuration, how we may further tune the beam polarizations in order to enhance (suppress) the signal (background) event cross sections and hence improve the sensitivity.

These results imply that if we further decrease $P_{e^+}$ to even $-100\%$ or increase $P_{e^-}$ to up to $+100\%$, we can boost the signal-process cross sections and lower the background-process cross sections simultaneously, thus improving the expected sensitivities.

\section{Projected sensitivity to the dark states with EM form factors}\label{sec:sens}

In this section, we will explain our event-selection procedure and present sensitivity reach.

\subsection{Event-selection procedure}\label{subsec:event_selection}

\begin{table}[t]
    \centering
    \begin{tabular}{cccc}
        \toprule
        \textbf{BP} & \textbf{BP1} & \textbf{BP2} & \textbf{BP3} \\ 
        \midrule
        \textbf{Definition} & $m_{\chi} = 5 \, \text{GeV}$ & $m_{\chi} = 50 \, \text{GeV}$ & $m_{\chi} = 110 \, \text{GeV}$ \\ 
        \textbf{Crit.~1} & $E_{\gamma} < 124.8 \, \text{GeV}$ & $E_{\gamma} < 105 \, \text{GeV}$ & $E_{\gamma} < 28.2 \, \text{GeV}$ \\ 
        \textbf{Crit.~2} & \multicolumn{3}{c}{$p_T^\gamma > 6 \, \text{GeV}, \quad |\eta_\gamma| < 2.79$} \\ 
        \textbf{Crit.~3} & \multicolumn{3}{c}{$E_\gamma < E_\gamma^Z - 5 \Gamma_\gamma^Z$ 
        } \\ 
        \bottomrule
        \toprule
        \textbf{BP} & \textbf{BP1} & \textbf{BP2} & \textbf{BP3} \\ 
        \midrule
        \textbf{Definition} & $m_{\chi} = 10 \, \text{GeV}$ & $m_{\chi} = 200 \, \text{GeV}$ & $m_{\chi} = 400 \, \text{GeV}$ \\ 
        \textbf{Crit.~1} & $E_{\gamma} < 499.8 \, \text{GeV}$ & $E_{\gamma} < 420 \, \text{GeV}$ & $E_{\gamma} < 180 \, \text{GeV}$ \\ 
        \textbf{Crit.~2} & \multicolumn{3}{c}{$p_T^\gamma > 6 \, \text{GeV}, \quad |\eta_\gamma| < 2.79$} \\ 
        \textbf{Crit.~3} & \multicolumn{3}{c}{$E_\gamma < E_\gamma^Z - 5 \Gamma_\gamma^Z$ 
        } \\ 
        \bottomrule
        \toprule
        \textbf{BP} & \textbf{BP1} & \textbf{BP2} & \textbf{BP3} \\ 
        \midrule
        \textbf{Definition} & $m_{\chi} = 100 \, \text{GeV}$ & $m_{\chi} = 600 \, \text{GeV}$ & $m_{\chi} = 1200 \, \text{GeV}$ \\ 
        \textbf{Crit.~1} & $E_{\gamma} < 1493.3 \, \text{GeV}$ & $E_{\gamma} < 1260 \, \text{GeV}$ & $E_{\gamma} < 540 \, \text{GeV}$ \\ 
        \textbf{Crit.~2} & \multicolumn{3}{c}{$p_T^\gamma > 60 \, \text{GeV}, \quad |\eta_\gamma| < 2.44$} \\ 
        \bottomrule
    \end{tabular}
    \caption{Event-selection criteria at different BPs for all the considered operator types at ILC and CLIC.
    The upper, middle, and lower parts of the table correspond to $\sqrt{s} = 250\,\text{GeV}$ at ILC, $1\,\text{TeV}$ at ILC, and $3\,\text{TeV}$ at CLIC, respectively.
    Crit.~3 shown in the upper and middle tables are effective for BP1 only.
    The background events are analyzed separately for each BP accordingly.
    $E_\gamma^Z \equiv (s - M_Z^2)/(2\sqrt{s})$ and $\Gamma_\gamma^Z \equiv M_Z \Gamma_Z/\sqrt{s}$.}
    \label{tab:ILC-CLIC-cut}
\end{table}

\begin{figure}[t]
    \centering
    \begin{subfigure}[t]{0.33\linewidth}
        \centering
        \includegraphics[width=\linewidth, height=6cm, keepaspectratio]{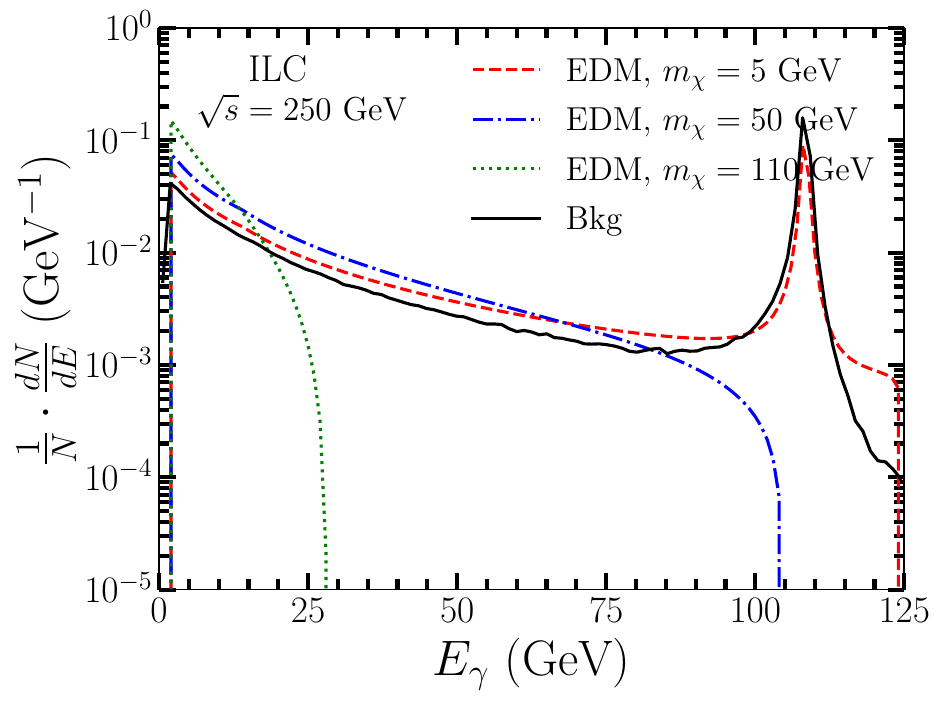}
        
        \vspace{0.5cm}
        
        \includegraphics[width=\linewidth, height=6cm, keepaspectratio]{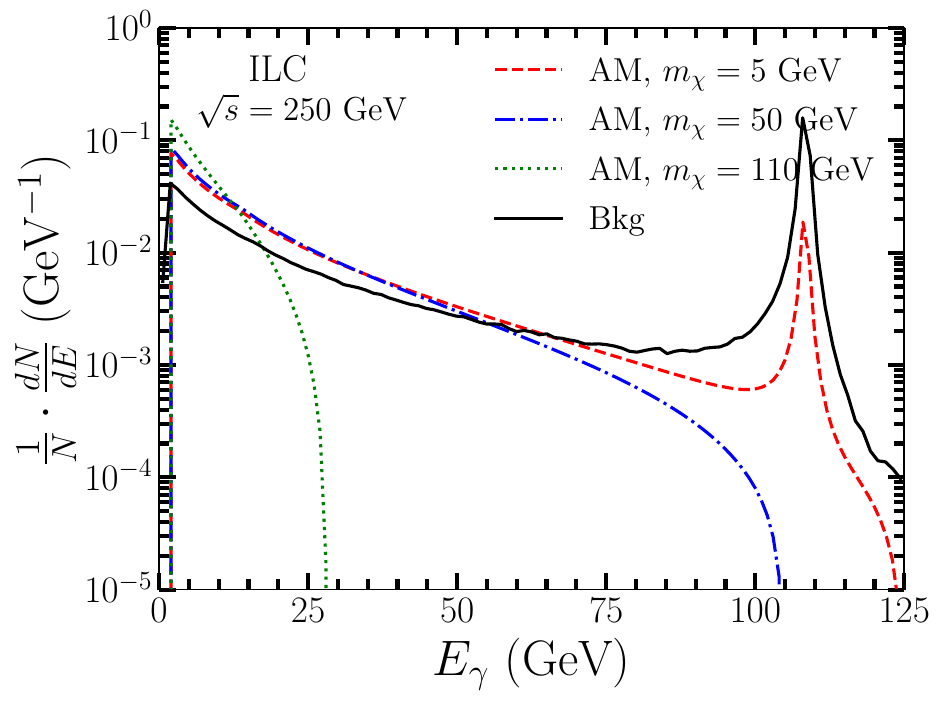}
    \end{subfigure}%
    \hfill
    \begin{subfigure}[t]{0.33\linewidth}
        \centering
        \includegraphics[width=\linewidth, height=6cm, keepaspectratio]{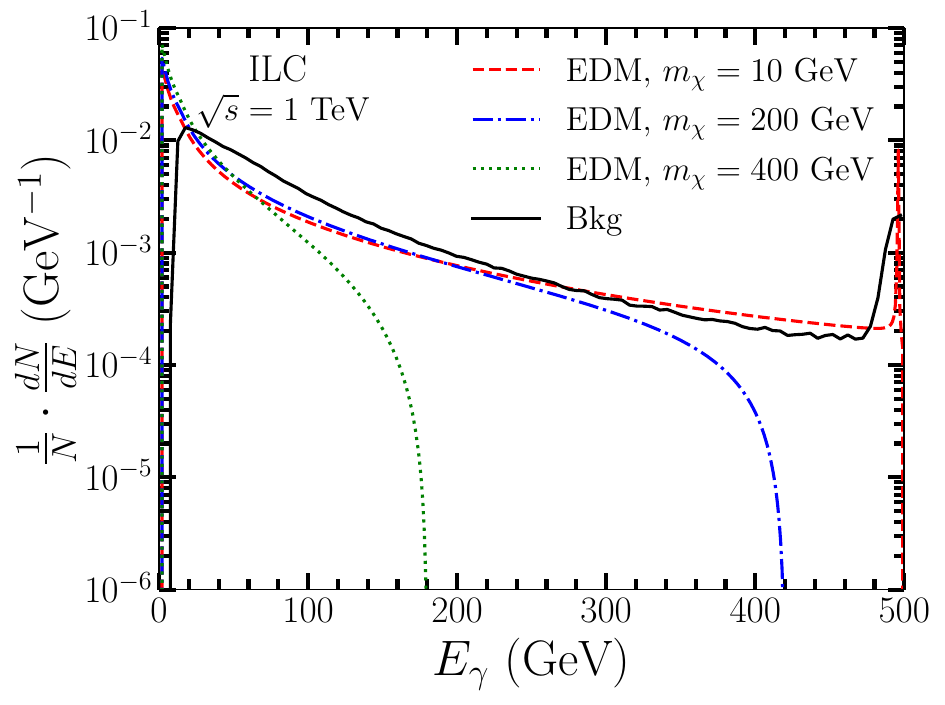}
        
        \vspace{0.5cm}
        
        \includegraphics[width=\linewidth, height=6cm, keepaspectratio]{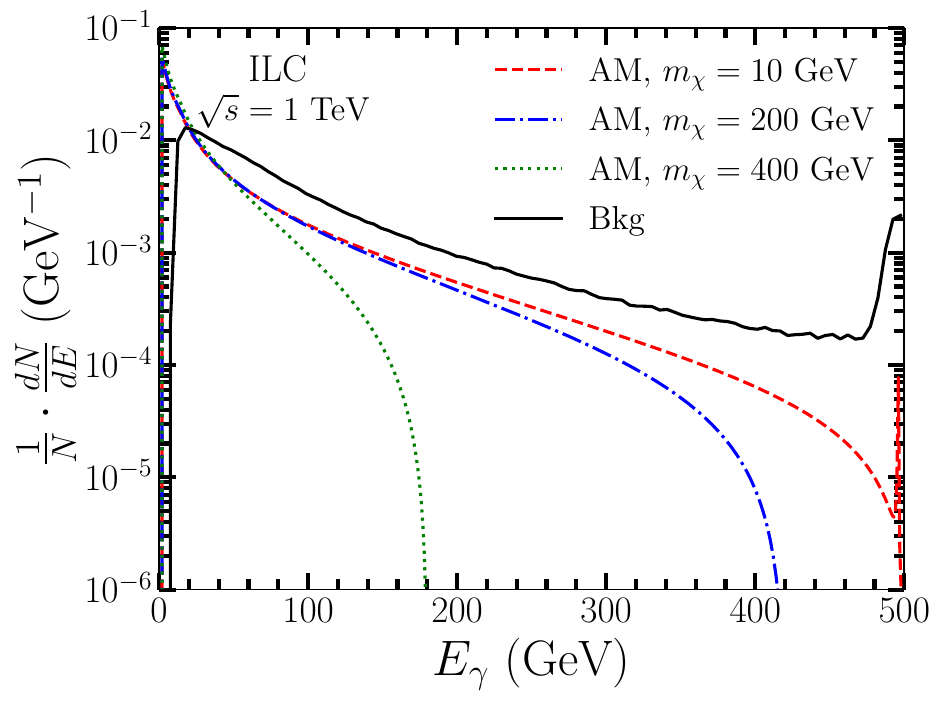}
    \end{subfigure}%
    \hfill
    \begin{subfigure}[t]{0.33\linewidth}
        \centering
        \includegraphics[width=\linewidth, height=6cm, keepaspectratio]{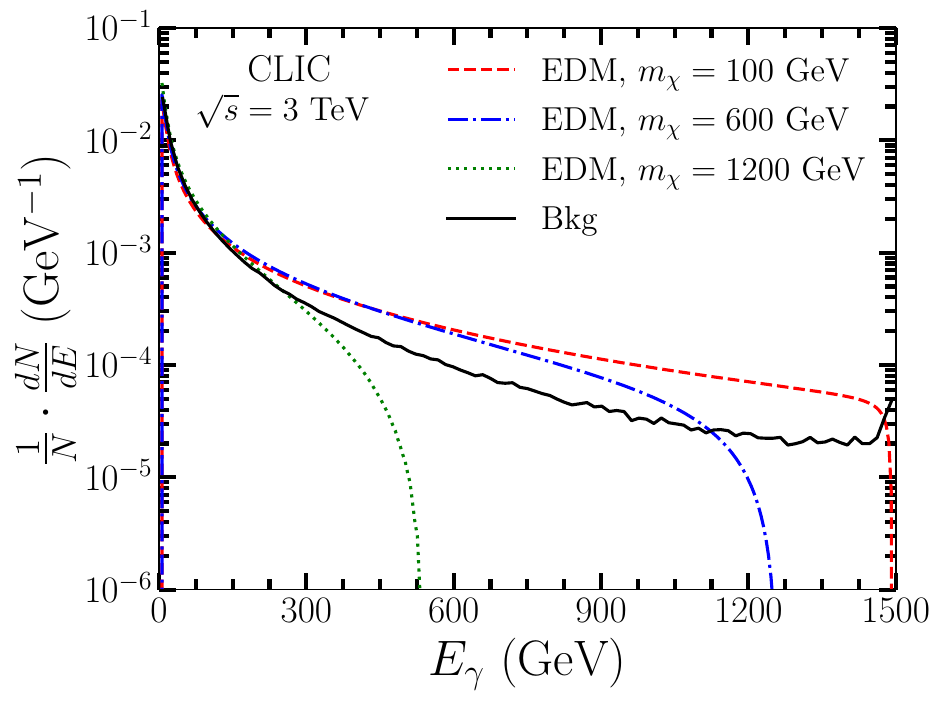}
        
        \vspace{0.5cm}
        
        \includegraphics[width=\linewidth, height=6cm, keepaspectratio]{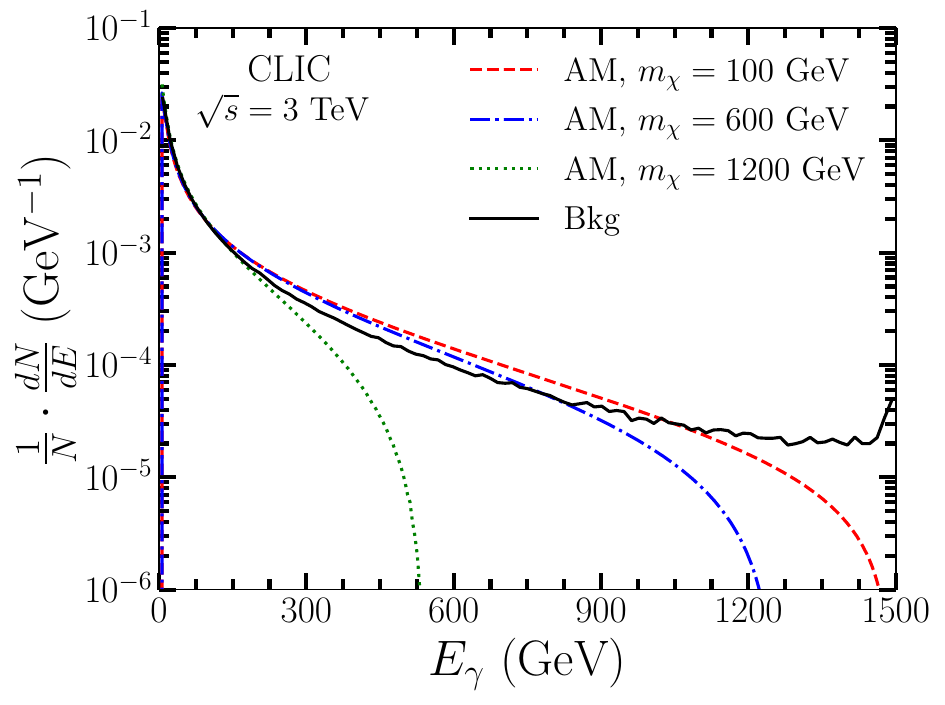}
    \end{subfigure}
\caption{Normalized $E_\gamma$-distributions at ILC with $\sqrt{s} = 250\,\text{GeV}$ (left column), ILC with $1\,\text{TeV}$ (middle column), and CLIC with $\sqrt{s} = 3\,\text{TeV}$ (right column), with the kinematic cuts given in equation~\eqref{eq:initial-cut} included.
The beams are assumed to be unpolarized.
The distributions are for both the irreducible SM background process $e^+e^- \rightarrow \bar{\nu}\nu\gamma$, and dark-state production process via dimension-5 MDM (top) and dimension-6 AM (bottom) operators.
For ILC with $\sqrt{s}=250$ GeV (1 TeV), three benchmark points with $m_\chi = 5, 50$, and $110$ GeV ($m_\chi = 10, 200, 400$ GeV) are chosen, while for CLIC with $\sqrt{s}=3$ TeV we choose $m_\chi = 100, 600$, and $1200$ GeV.
}
\label{fig:Egamma_ILC_CLIC}
\end{figure}

We select various benchmark points (BPs) of $m_\chi$ for ILC and CLIC, summarized in table~\ref{tab:ILC-CLIC-cut}.  
We show in figure~\ref{fig:Egamma_ILC_CLIC} the normalized distribution of the visible photon's energy, $E_\gamma$, in both signal and background processes, at ILC with $\sqrt{s} = 250\,\text{GeV}$ and $1\,\text{TeV}$, and at CLIC with $\sqrt{s} = 3\,\text{TeV}$, with the cuts listed in equation~\eqref{eq:initial-cut} included.  
In detail, these distributions entail both the irreducible SM background process $e^-e^+ \rightarrow \nu \bar{\nu} \gamma$ and the $\chi$-production process $e^- e^+ \to \chi \bar{\chi} \gamma$ mediated by the dimension-5 EDM and dimension-6 AM operators.
We note that here we assume no polarization for both the incoming electron and positron beams.

For ILC with $\sqrt{s} = 250\,\text{GeV}$ (corresponding to the first column of figure~\ref{fig:Egamma_ILC_CLIC}), the event rates for both EDM and AM signals drop sharply when the photon energy approaches its kinematic threshold $E_\gamma^{\text{max}} = \frac{s - 4m_\chi^2}{2\sqrt{s}}$, while the distribution of the SM background process remains relatively stable across the shown range of $E_\gamma$ except for the peak observed at $E_\gamma^Z\equiv\frac{s-m_Z^2}{2\sqrt{s}}$ where $m_Z$ is the $Z$-boson mass (see more detailed discussion below).
The kinematic thresholds $E_\gamma^{\text{max}}$ are computed to be 124.8 GeV, 105 GeV, and 28.2 GeV for $m_\chi = 5\,\text{GeV}$, $50\,\text{GeV}$, and $110\,\text{GeV}$, respectively.
Selecting events with $E_\gamma$ below these respective energy thresholds allows for good separation between signal and background events (not so good, though, for the case of $m_\chi=5$ GeV which is below $m_Z/2$ where a peak at $E_\gamma^Z$ appears; see more detail in the last paragraph of this subsection), as can be seen in figure~\ref{fig:Egamma_ILC_CLIC}, forming the basis for Crit.~1 in the upper part of table~\ref{tab:ILC-CLIC-cut}.

A similar pattern is observed in the ILC scenario with $\sqrt{s} = 1\,\text{TeV}$ (the second column of figure~\ref{fig:Egamma_ILC_CLIC}) as well as in the CLIC case (the third column of figure~\ref{fig:Egamma_ILC_CLIC}).
We thus choose Crit.~1 in the middle and lower tables of table~\ref{tab:ILC-CLIC-cut} also with $E_\gamma < \frac{s - 4m_\chi^2}{2\sqrt{s}}$, in order to enhance the discrimination between signal and background events.

At ILC~\cite{Habermehl:2020njb, Habermehl:2018yul}, the BeamCal configuration imposes two distinct cuts on $p_{T}^{\gamma}$: $p_{T}^{\gamma} > 1.92$ GeV and $p_{T}^{\gamma} > 5.65$ GeV, depending on the photon's azimuthal angle.
For simplicity, we adopt a universal, conservative $p_{T}^{\gamma}$ cut, i.e.~$p_{T}^{\gamma} > 6$ GeV.
Additionally, the signal significance is influenced by the photon-reconstruction efficiency.
For photons satisfying $E_\gamma > 2$ GeV and $7^\circ < \theta_\gamma < 173^\circ$ (or equivalently $|\eta_\gamma| < 2.79$), the reconstruction efficiency at ILC exceeds $99\%$~\cite{Habermehl:2018yul}.
Therefore, we impose a cut on $\eta_\gamma$, $|\eta_\gamma|<2.79$~\cite{Habermehl:2020njb}.
These cuts are given as Crit.~2 in the upper and middle tables of table~\ref{tab:ILC-CLIC-cut}.
We impose similar cuts at CLIC, following Ref.~\cite{Blaising:2021vhh}, given as Crit.~2 in the lower part of table~\ref{tab:ILC-CLIC-cut}.

In the first two columns of figure~\ref{fig:Egamma_ILC_CLIC}, each panel exhibits two distinct resonance peaks.
At ILC with $\sqrt{s} = 250\,\text{GeV}$ (left column), the two peaks appear around $E_\gamma \sim E_\gamma^Z \approx 108\,\text{GeV}$ — one from the SM background and the other from the signal process with $m_\chi = 5\,\text{GeV}$.
Similarly, at ILC with $\sqrt{s} = 1\,\text{TeV}$ (middle column), two resonance peaks are observed around $E_\gamma \sim E_\gamma^Z \approx 495\,\text{GeV}$, corresponding to the background and the signal with $m_\chi = 10\,\text{GeV}$.
These resonance peaks at $E_\gamma^Z$ arise from
the resonant decay of the $Z$-boson into a pair of fermions — neutrinos in the background process and dark fermions $\chi\bar{\chi}$ in the signal process — since both the neutrino mass and $m_\chi = 5\,\text{GeV}$ or $10\,\text{GeV}$ are below the kinematic threshold $m_Z/2$.
For these resonances around $E_\gamma^Z$, the full-width-at-half-maximum of the background resonance is given by $\Gamma_\gamma^Z \equiv M_Z \Gamma_Z/\sqrt{s}$ where $\Gamma_Z$ is the $Z$-boson width.
For the other ILC benchmark points and for all the CLIC benchmark points, we have $m_\chi>m_Z/2$, and the resonance feature is thus absent.
Correspondingly, we impose Crit.~3, $E_\gamma < E_\gamma^Z - 5 \Gamma_\gamma^Z$, in the upper and middle parts of table~\ref{tab:ILC-CLIC-cut}, effective for the BP1 cases at ILC only.

\subsection{Numerical results}\label{subsec:results}

After applying all the event-selection criteria, we proceed to evaluate the signal significance using the following definition:
\begin{equation}
\chi^2(\mathcal{C}_\chi) \equiv \frac{S^2(\mathcal{C}_\chi)}{S(\mathcal{C}_\chi) + B + (\epsilon B)^2},
\end{equation}
where $S$ and $B$ represent the number of signal and background events, respectively, for a given integrated luminosity, $\mathcal{C}_\chi$ is defined in section~\ref{sec:model}, and $\epsilon$ is the corresponding systematic uncertainty in the background estimation.
By calculating $\chi^2(C_\chi) - \chi^2(0) = 2.71$, we can determine the $95\%$ confidence level (C.L.) upper limits on the EM form factors of $\chi$.

\begin{figure}[t]
\centering
\includegraphics[width=0.495\textwidth]{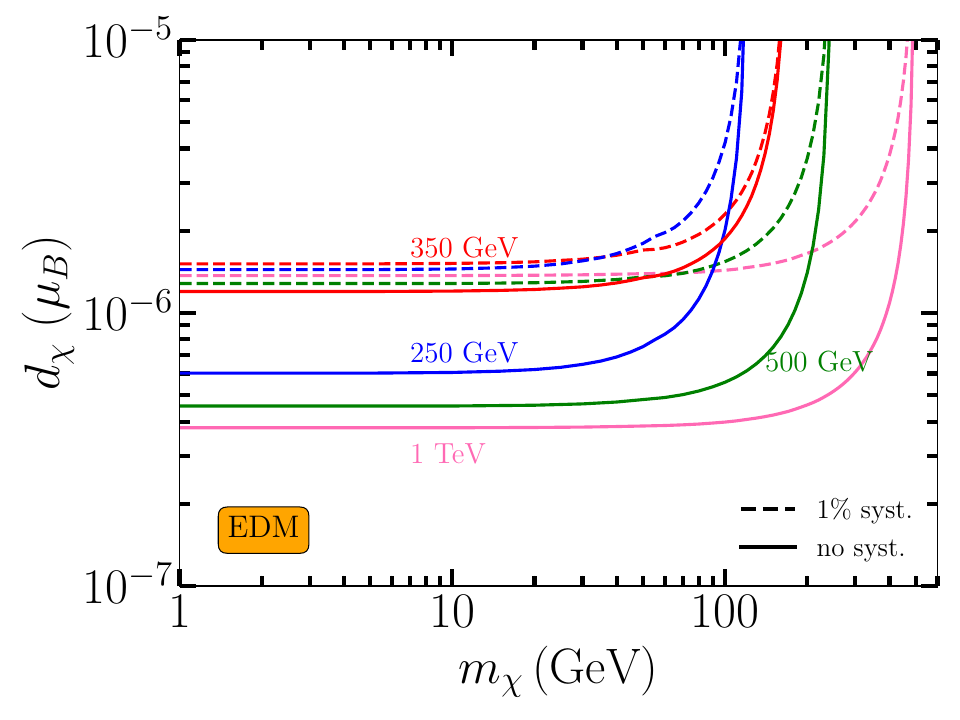}
\includegraphics[width=0.495\textwidth]{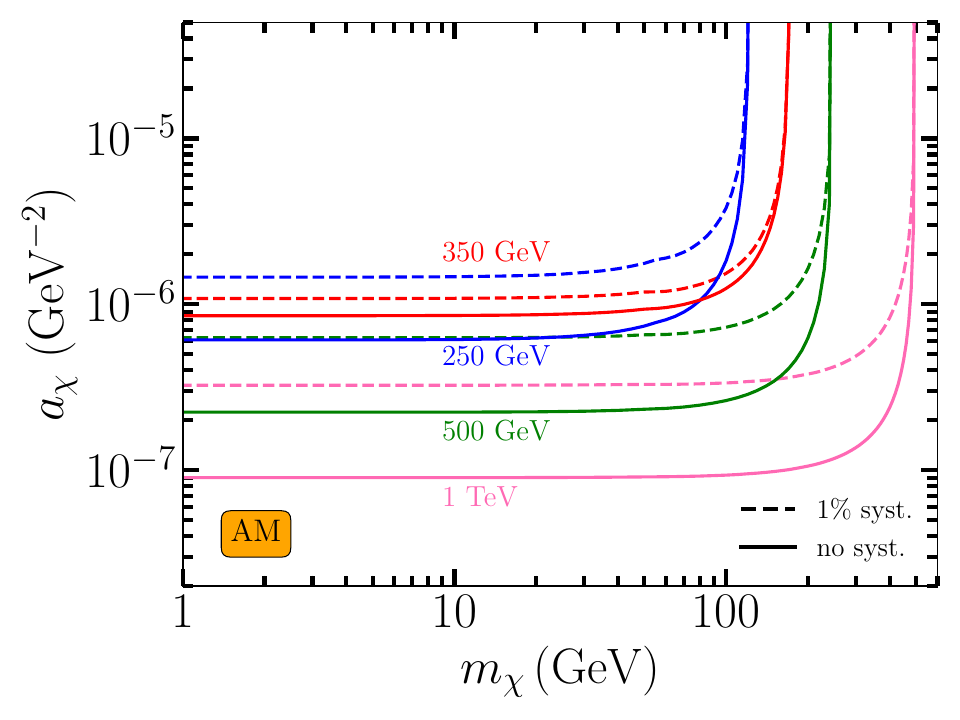}
\caption{The expected 95\% C.L.~upper limits on the EM form factors at ILC for the dimension-5 operator EDM (left) and for the dimension-6 operator AM (right), under optimal polarization configurations.
The projected limits are presented at four representative COM energies: $\sqrt{s} = 250\, \text{GeV}$ (blue), $350\,\text{GeV}$ (red), $500\,\text{GeV}$ (green), and $1\,\text{TeV}$ (pink).
For $\sqrt{s} = 1\,\text{TeV}$, the polarization configuration $(P_{e^-}, P_{e^+}) = (+80\%, -20\%)$ is used, while for the other three COM energies, $(P_{e^-}, P_{e^+}) = (+80\%, -30\%)$ is applied.
The solid (dashed) lines correspond to scenarios with zero ($1\%$) background systematic uncertainty.}

\label{fig:ILC_Limits_Uncertainty}
\end{figure}

In figure~\ref{fig:ILC_Limits_Uncertainty}, we present the projected upper bounds on the EM form factors at ILC under the optimal polarization conditions: $(P_{e^-}, P_{e^+}) = (+80\%, -30\%)$ for $\sqrt{s} = 250, 350,$ and $500\,\text{GeV}$, and $(+80\%, -20\%)$ for $\sqrt{s} = 1000\,\text{GeV}$. The corresponding integrated luminosities are $0.9, 0.044, 1.6,$ and $3.2$ ab$^{-1}$, respectively.
The solid (dashed) lines represent scenarios with zero $(1\%)$ background systematic uncertainty. We observe that for the EDM and AM at $\sqrt{s} = 1000\, \text{GeV}$ with an integrated luminosity of 3.2 ab$^{-1}$, ILC can constrain the coupling coefficients $d_\chi$ of the dimension-5 operator to approximately $3.80 \times 10^{-7}\, \mu_B$ ($1.37 \times 10^{-6}\, \mu_B$) and $a_\chi$ of the dimension-6 operator to approximately $9.01 \times 10^{-8}\, \text{GeV}^{-2}$ ($3.25 \times 10^{-7}\, \text{GeV}^{-2}$), corresponding to the background systematic uncertainty of $0 (1\%)$, respectively. Clearly, under ideal conditions with vanishing background systematic uncertainty, the sensitivity of ILC to the EM form factors of the dark states improves by a factor of $\sim 3$ compared to the case with a systematic uncertainty of $1\%$. Therefore, precise control over the background systematic uncertainties is crucial for improving sensitivity reach.

\begin{figure}[t]
\centering
\includegraphics[width=0.495\textwidth]{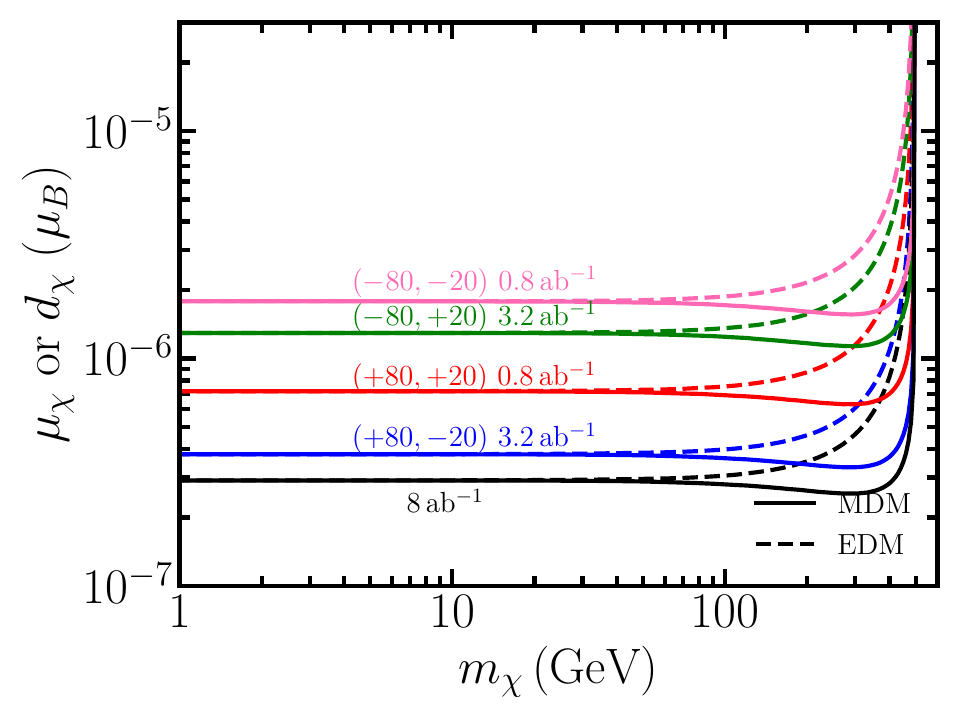}
\includegraphics[width=0.495\textwidth]{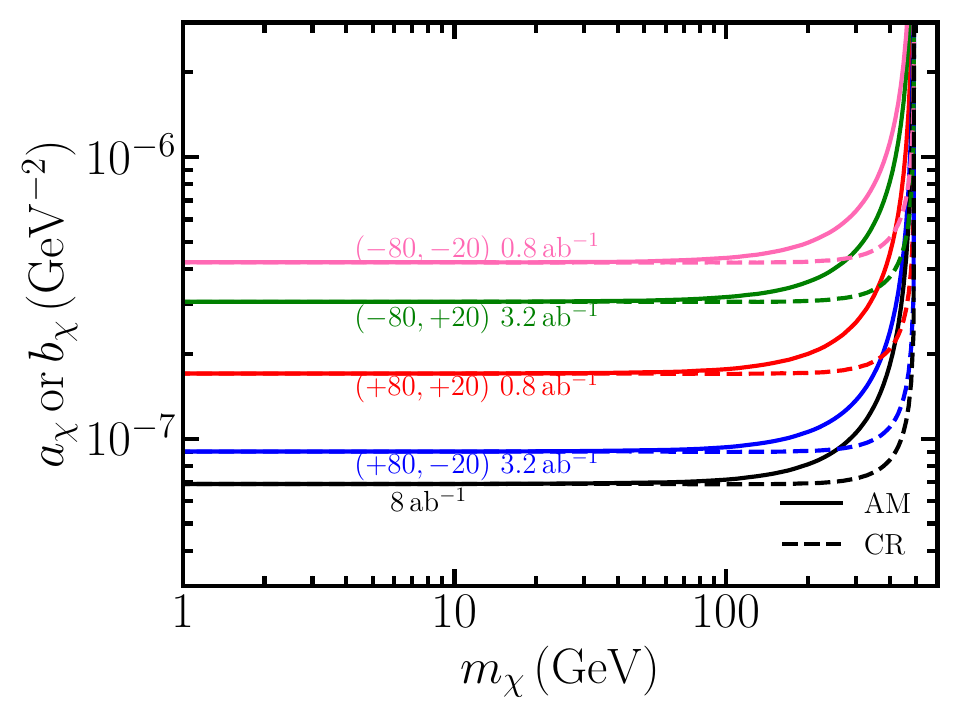}
\caption{The expected 95\% C.L.~upper limits on the EM form factors at ILC for $\sqrt{s} = 1 \, \text{TeV}$, considering different polarization modes.
The left panel shows limits for dimension-5 operators: MDM (solid) and EDM (dashed), while the right panel corresponds to dimension-6 operators: AM (solid) and CR (dashed).
The expected limits are represented with the following color scheme: red for $P(e^-, e^+) = (+80\%, +20\%)$ with $0.8 \, \text{ab}^{-1}$; black for $P(e^-, e^+) = (-80\%, -20\%)$ with $0.8 \, \text{ab}^{-1}$; green for $P(e^-, e^+) = (-80\%, +20\%)$ with $3.2 \, \text{ab}^{-1}$; cyan for $P(e^-, e^+) = (+80\%, -20\%)$ with $3.2 \, \text{ab}^{-1}$; and blue for the combined total luminosity of $8 \, \text{ab}^{-1}$.}
\label{fig:ILC_Limits_polarization}
\end{figure}

\begin{figure}[t]
\centering
\includegraphics[width=0.495\textwidth]{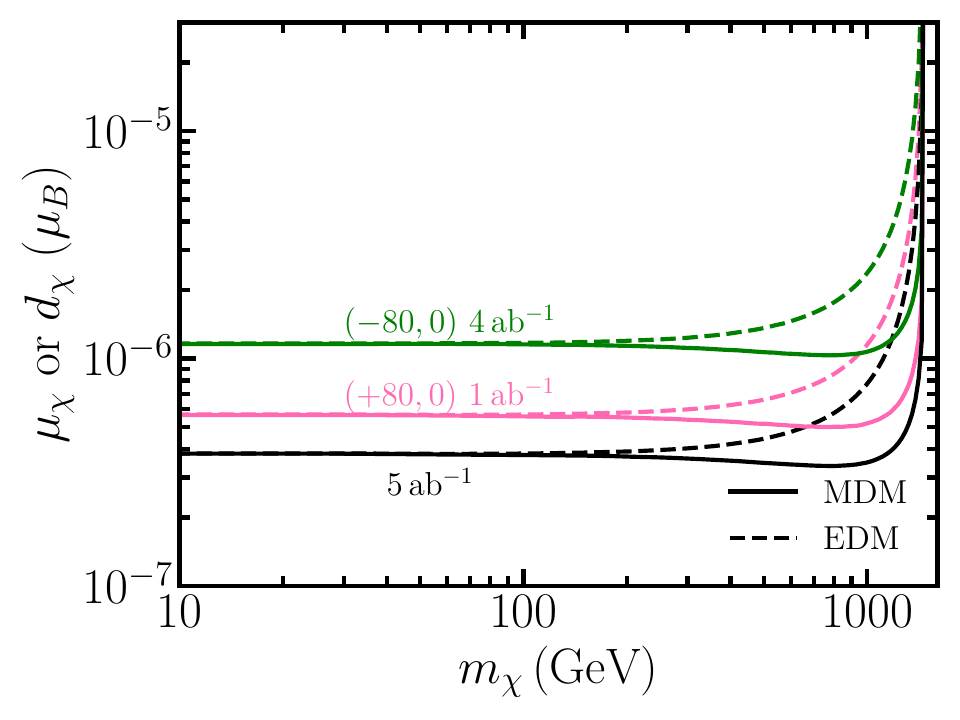}
\includegraphics[width=0.495\textwidth]{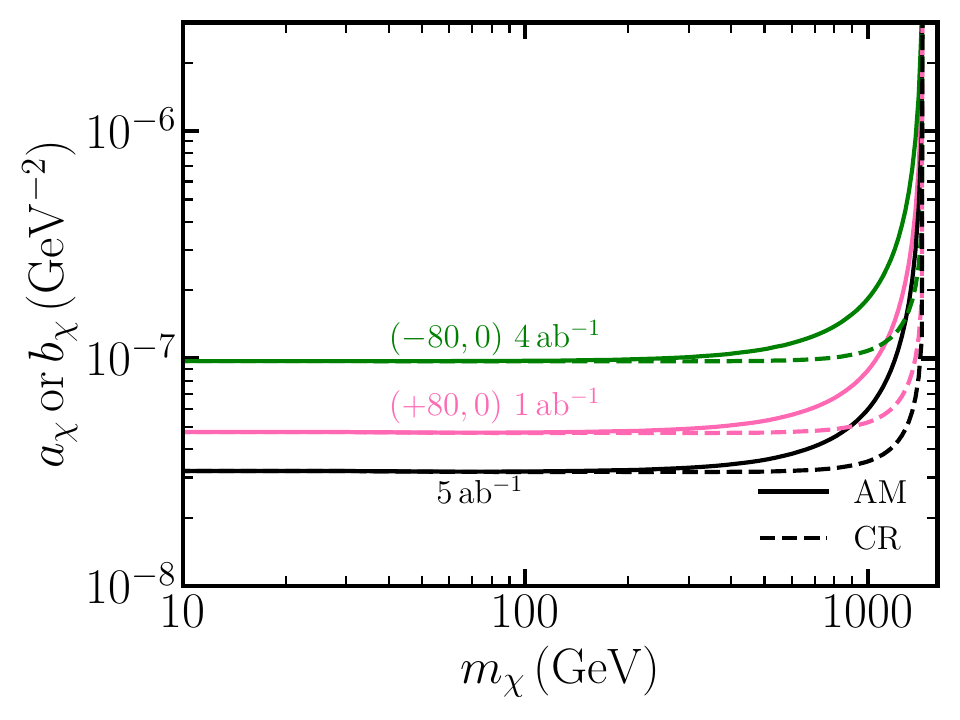}
\caption{The expected 95\% C.L.~upper limits on the EM form factors at CLIC for $\sqrt{s} = 3 \, \text{TeV}$, considering different polarization modes.
The left panel shows limits for dimension-5 operators: MDM (solid) and EDM (dashed), while the right panel corresponds to dimension-6 operators: AM (solid) and CR (dashed).
The expected limits are represented as follows: red for $P(e^-, e^+) = (-80\%, 0\%)$ with $4 \, \text{ab}^{-1}$; black for $P(e^-, e^+) = (+80\%, 0\%)$ with $1\,\text{ab}^{-1}$; and blue for the total luminosity of $5\, \text{ab}^{-1}$ combining both running modes.}
\label{fig:CLIC_Limits_polarization}
\end{figure}

To further enhance the sensitivity, we perform a weighted combination of all the polarization configurations at ILC with $\sqrt{s} = 1\,\text{TeV}$, defined as
\begin{equation}
\chi_c^2 = \sum\limits_{i} \chi_i^2 = \sum\limits_{i} \frac{S_i^2}{S_i + B_i + (\epsilon B_i)^2},
\end{equation}
where the index $i$ runs over all the polarization settings.
The results in sensitivity reach with a combined integrated luminosity of $8\,\text{ab}^{-1}$, shown in figure~\ref{fig:ILC_Limits_polarization}.

We find in these results that ILC can constrain the coupling coefficients of dimension-5 operators to $2.9 \times 10^{-7}\, \mu_B$ and the coupling coefficients of dimension-6 operators to $6.9 \times 10^{-8}\, \text{GeV}^{-2}$. Compared with the case where we include only the optimal polarization configuration $P(e^-, e^+) = (80\%, -20\%)$, this approach achieves improvement on the sensitivities under the same conditions. It is worth noting that for $m_\chi \gtrsim 50\,\text{GeV}$, the constraint on $d_\chi$ deteriorates quickly with increasing $m_\chi$ while that on $\mu_\chi$ gets even strengthened slightly before turning negligible sharply right below the kinematic thresholds. These behaviors approximately match with those of the corresponding signal cross sections illustrated in figure~\ref{fig:sig-mchi-1tev}.

Similarly, figure~\ref{fig:CLIC_Limits_polarization} illustrates the 95\% C.L.~upper limits on the EM form factors of the dark-state $\chi$ at CLIC with $\sqrt{s} = 3$~TeV, for two polarization configurations as well as their weighted combinations.
The figure once again demonstrates that the weighted combinations can further enhance the sensitivity.
For example, for the dimension-6 operators $a_\chi$ and $b_\chi$, the weighted combination improves the sensitivity by approximately $67\%$ compared to the single polarization configuration $P(e^-, e^+) = (-80\%, 0\%)$ with an integrated luminosity of 4 ab$^{-1}$.
We also note that CLIC with the polarization scheme $P(e^-, e^+) = (+80\%, 0\%)$ is shown to have stronger sensitivities than that with $P(e^-, e^+) = (-80\%, 0\%)$, even though the former scheme is planned to have an integrated luminosity only one quarter of that of the latter scheme; this is mainly due to the larger signal-to-background cross-section ratio with the scheme $P(e^-, e^+) = (+80\%, 0\%)$, cf.~table~\ref{table:difpol-SM-sig-CLIC}.

\begin{figure}[t]
    \includegraphics[width=0.495\textwidth]{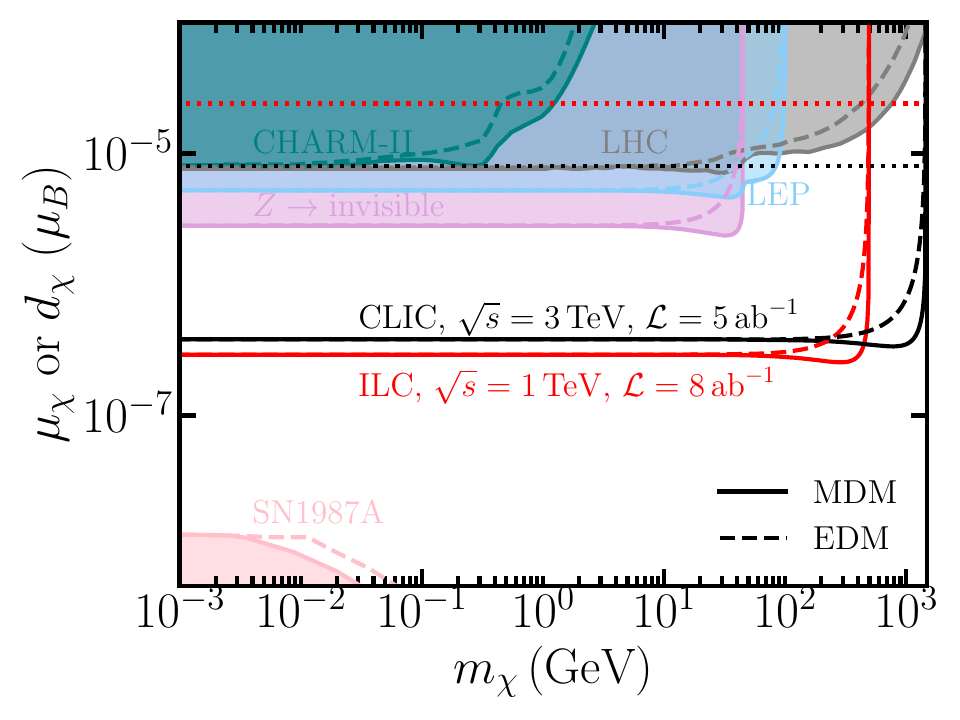}
    \includegraphics[width=0.495\textwidth]{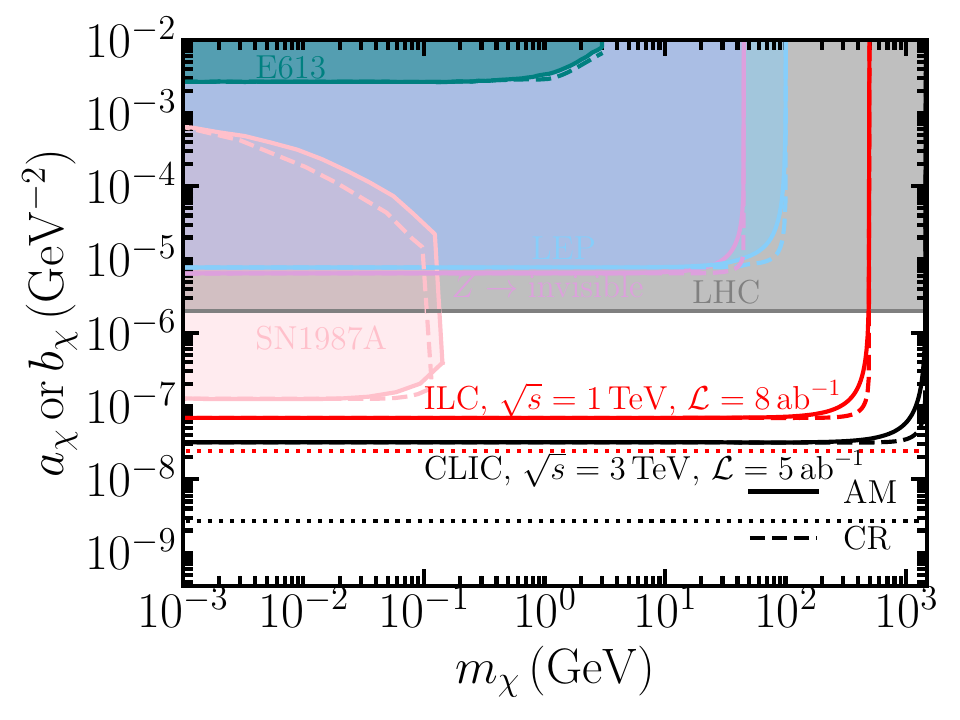}
    \caption{The expected 95\% C.L.~upper limits on the EM form factors at ILC and CLIC, for dimension-5 operators (left) through MDM (solid) and EDM (dashed) interactions, and for dimension-6 operators (right) through AM (solid) and CR (dashed) interactions, all under total polarization conditions and various polarization modes.
    The expected bounds are plotted with the following color scheme: red for ILC with $\sqrt{s} = 1 \, \text{TeV}$ and an integrated luminosity of $8 \, \text{ab}^{-1}$; and black for CLIC with $\sqrt{s} = 3 \, \text{TeV}$ and an integrated luminosity of $5 \, \text{ab}^{-1}$.
    The red and black dotted lines correspond to $\Lambda=1\text{~TeV}$ and $\Lambda=3\text{~TeV}$ for $C_i^\gamma= 4\pi$ (see equation~\eqref{eq:WC_loop} and its relevant discussion), where 1~TeV and 3~TeV are the COM energies at ILC and CLIC, respectively.
    For larger values of $C_\chi$, the sensitivity estimates become less reliable as a result of the breakdown of perturbative unitarity.
}
    \label{fig:CLIC_ILC_Limit}
\end{figure}

Finally, in figure~\ref{fig:CLIC_ILC_Limit} we present the $95\%$ C.L.~exclusion bounds on the considered EM form factors, at ILC with $\sqrt{s} = 1$~TeV and CLIC with $\sqrt{s} = 3$~TeV, as functions of $m_\chi$, combining different polarization configurations with the corresponding integrated luminosities.
For dimension-5 operators (left panel), the constraints are given via MDM (solid line) and EDM (dashed line) interactions, while for dimension-6 operators (right panel), they are represented through AM (solid line) and CR (dashed line) interactions.

Existing constraints are shown as shaded regions.
These existing bounds originate from both terrestrial experiments and astrophysical observations.
The terrestrial experiments include proton beam-dump experiments CHARM-II and E613~\cite{Chu:2020ysb}, LEP~\cite{Zhang:2022ijx} with mono-photon searches and searches for $Z$-boson invisible decays, and the LHC~\cite{Arina:2020mxo} with monojet searches.
The astrophysical bounds are derived from the observation of the supernova SN1987A~\cite{Chu:2019rok}.
It is important to note that only the most competitive constraints are displayed in figure~\ref{fig:CLIC_ILC_Limit}, while a more comprehensive summary of the constraints can be found in, e.g.~Refs.~\cite{Chu:2018qrm,Chu:2020ysb,Chu:2019rok}.

In figure~\ref{fig:CLIC_ILC_Limit}, we observe that our projected sensitivity reach exceeds the existing bounds by about 1--2 orders of magnitude depending on the type of the EM form factor and the mass of $\chi$. For instance, the EDM sensitivity at CLIC with $\sqrt{s}=3$ TeV and $\mathcal{L}=5$ ab$^{-1}$ can be as low as $3.82\times 10^{-7}\, \mu_B$ in most of the accessible mass range, while the AM sensitivity goes down to even $3.2\times 10^{-8}\, \text{GeV}^{-2}$. Moreover, comparing the two panels of figure~\ref{fig:CLIC_ILC_Limit}, we observe that the sensitivities to the dimension-5 operators are better at ILC ($\sqrt{s} = 1$~TeV, with an integrated luminosity of 8~ab$^{-1}$) than at CLIC ($\sqrt{s} = 3$~TeV, with 5~ab$^{-1}$). In contrast, for the dimension-6 operators, the sensitivity comparison between the two colliders is reverted. This difference arises because the signal cross sections induced by the dimension-6 operators increase more rapidly with the COM energy $\sqrt{s}$ than those from the dimension-5 operators, cf.~figure~\ref{fig:sig-s-m}.

In addition, based on the discussion on perturbative unitarity in section~\ref{sec:model}, we note the importance of the validity of the effective-field-theory (EFT) approach we have taken~\cite{Barducci:2025twe}, which breaks down if the probed new-physics scales fall below the COM energy of the collider,  $\Lambda\lesssim \sqrt{s}$.
Accordingly, in figure~\ref{fig:CLIC_ILC_Limit}, we show red and black dotted lines, corresponding to $\Lambda=1\text{~TeV}$ and $\Lambda = 3\text{~TeV}$, for $C_i^\gamma= 4\pi$, respectively. 
The results indicate that for the dimension-5 operators, the EFT remains valid across almost the whole sensitive range of $m_\chi$.
However, for the dimension-6 operators, the probed new-physics scales are below the collider COM energies by approximately a factor of 3--10, thus rendering the EFT validity break down.
We emphasize that these dotted lines are not strict bounds but should be interpreted only as a reference at the order-of-magnitude level; for values of $C_\chi$ above these dotted lines, the sensitivity estimates become less reliable owing to loss of perturbative unitarity or EFT validity.

\section{Conclusions}\label{sec:conclu}

In this study, we systematically explore the sensitivity of future linear electron-positron colliders (ILC and CLIC) to dark-state EM form factors as functions of the mass of $\chi$, focusing on the MDM and EDM at dimension 5, and the AM and CR at dimension 6.
Further, we have highlighted the feasibility to improve the sensitivities by tuning the incoming beams' polarizations at ILC and CLIC.
In particular, we find that for CLIC with $\sqrt{s}=3$ TeV, the polarization scheme $P(e^-, e^+) = (+80\%, 0\%)$ enhances (suppresses) the signal (background) cross sections simultaneously compared to the scheme $P(e^-, e^+) = (-80\%, 0\%)$; further, the former polarization scheme is estimated to provide stronger final constraining power on the dark-state EM form factors despite its relatively less assigned operation time.
On the other hand, for ILC with $\sqrt{s}=1$ TeV, the polarization mode with the optimal signal-to-background cross-section ratio is entertained with the most operation time [$40\%$ operation time assigned for the polarization scheme, $P(e^-, e^+) = (+80\%, -20\%)$], allowing for large enhancement on the ILC sensitivities.

We have simulated both signal and background processes with MC techniques, and imposed appropriate selection cuts.
Our numerical results show that suppressing systematic uncertainties to zero can enhance the sensitivity to the EM form factors by a factor up to about 3, compared to a benchmark of the systematic uncertainty of 1\%.
In the final sensitivity results, we find that considering operation conditions of ILC ($\sqrt{s} = 1\, \text{TeV},\, \mathcal{L}_{\text{int}}=8\, \text{ab}^{-1}$) and CLIC ($\sqrt{s} = 3\, \text{TeV},\, \mathcal{L}_{\text{int}}=5\, \text{ab}^{-1}$), the EM form factors of dark-sector particles can be constrained to levels about one to two orders of magnitude below the current experimental limits.

Finally, we note that the validity of the EFT is retained only if the probed new-physics cutoff scale $\Lambda$ is higher than the COM energy.
We have performed order-of-magnitude estimates which show that for the dimension-5 operators ILC and CLIC can probe the corresponding form factors well below the levels required by both perturbative unitarity and EFT validity.
However, for the dimension-6 operators, the values of $a_\chi$ and $b_\chi$ to which ILC and CLIC are sensitive are above the levels determined by the above-mentioned theoretical requirements by a factor of $\sim 3$--$10$, thus rendering the perturbative computation less reliable.

\acknowledgments
This work was supported by the National Natural Science Foundation of China under grant Nos.~12475106 and 12505120, and the Fundamental Research Funds for the Central Universities under Grant No.~JZ2025HGTG0252.
Z.~S.~Wang would like to thank the Particle Theory and Cosmology Group of IBS-CTPU, Daejeon, South Korea, for their hospitality, where part of this work was completed.

\bibliographystyle{JHEP}
\bibliography{references.bib}

\providecommand{\href}[2]{#2}\begingroup\raggedright\begin{thebibliography}{10}

\bibitem{Antel:2023hkf}
C.~Antel et~al., \emph{{Feebly-interacting particles: FIPs 2022 Workshop
  Report}}, \href{http://dx.doi.org/10.1140/epjc/s10052-023-12168-5}{\emph{Eur.
  Phys. J. C} {\bf 83} (2023) 1122},
  [\href{http://arxiv.org/abs/2305.01715}{{\tt 2305.01715}}].

\bibitem{Pospelov:2000bq}
M.~Pospelov and T.~ter Veldhuis, \emph{{Direct and indirect limits on the
  electromagnetic form-factors of WIMPs}},
  \href{http://dx.doi.org/10.1016/S0370-2693(00)00358-0}{\emph{Phys. Lett. B}
  {\bf 480} (2000) 181--186}, [\href{http://arxiv.org/abs/hep-ph/0003010}{{\tt
  hep-ph/0003010}}].

\bibitem{Chu:2018qrm}
X.~Chu, J.~Pradler and L.~Semmelrock, \emph{{Light dark states with
  electromagnetic form factors}},
  \href{http://dx.doi.org/10.1103/PhysRevD.99.015040}{\emph{Phys. Rev. D} {\bf
  99} (2019) 015040}, [\href{http://arxiv.org/abs/1811.04095}{{\tt
  1811.04095}}].

\bibitem{DelNobile:2012tx}
E.~Del~Nobile, C.~Kouvaris, P.~Panci, F.~Sannino and J.~Virkajarvi,
  \emph{{Light Magnetic Dark Matter in Direct Detection Searches}},
  \href{http://dx.doi.org/10.1088/1475-7516/2012/08/010}{\emph{JCAP} {\bf 08}
  (2012) 010}, [\href{http://arxiv.org/abs/1203.6652}{{\tt 1203.6652}}].

\bibitem{Kavanagh:2018xeh}
B.~J. Kavanagh, P.~Panci and R.~Ziegler, \emph{{Faint Light from Dark Matter:
  Classifying and Constraining Dark Matter-Photon Effective Operators}},
  \href{http://dx.doi.org/10.1007/JHEP04(2019)089}{\emph{JHEP} {\bf 04} (2019)
  089}, [\href{http://arxiv.org/abs/1810.00033}{{\tt 1810.00033}}].

\bibitem{Arina:2020mxo}
C.~Arina, A.~Cheek, K.~Mimasu and L.~Pagani, \emph{{Light and Darkness:
  consistently coupling dark matter to photons via effective operators}},
  \href{http://dx.doi.org/10.1140/epjc/s10052-021-09010-1}{\emph{Eur. Phys. J.
  C} {\bf 81} (2021) 223}, [\href{http://arxiv.org/abs/2005.12789}{{\tt
  2005.12789}}].

\bibitem{NA64:2017vtt}
{\scshape NA64} collaboration, D.~Banerjee et~al., \emph{{Search for vector
  mediator of Dark Matter production in invisible decay mode}},
  \href{http://dx.doi.org/10.1103/PhysRevD.97.072002}{\emph{Phys. Rev. D} {\bf
  97} (2018) 072002}, [\href{http://arxiv.org/abs/1710.00971}{{\tt
  1710.00971}}].

\bibitem{Prinz:1998ua}
A.~A. Prinz et~al., \emph{{Search for millicharged particles at SLAC}},
  \href{http://dx.doi.org/10.1103/PhysRevLett.81.1175}{\emph{Phys. Rev. Lett.}
  {\bf 81} (1998) 1175--1178}, [\href{http://arxiv.org/abs/hep-ex/9804008}{{\tt
  hep-ex/9804008}}].

\bibitem{LDMX:2018cma}
{\scshape LDMX} collaboration, T.~{\r{A}}kesson et~al., \emph{{Light Dark
  Matter eXperiment (LDMX)}},  \href{http://arxiv.org/abs/1808.05219}{{\tt
  1808.05219}}.

\bibitem{BDX:2016akw}
{\scshape BDX} collaboration, M.~Battaglieri et~al., \emph{{Dark Matter Search
  in a Beam-Dump eXperiment (BDX) at Jefferson Lab}},
  \href{http://arxiv.org/abs/1607.01390}{{\tt 1607.01390}}.

\bibitem{Chu:2020ysb}
X.~Chu, J.-L. Kuo and J.~Pradler, \emph{{Dark sector-photon interactions in
  proton-beam experiments}},
  \href{http://dx.doi.org/10.1103/PhysRevD.101.075035}{\emph{Phys. Rev. D} {\bf
  101} (2020) 075035}, [\href{http://arxiv.org/abs/2001.06042}{{\tt
  2001.06042}}].

\bibitem{LSND:1996jxj}
{\scshape LSND} collaboration, C.~Athanassopoulos et~al., \emph{{The Liquid
  scintillator neutrino detector and LAMPF neutrino source}},
  \href{http://dx.doi.org/10.1016/S0168-9002(96)01155-2}{\emph{Nucl. Instrum.
  Meth. A} {\bf 388} (1997) 149--172},
  [\href{http://arxiv.org/abs/nucl-ex/9605002}{{\tt nucl-ex/9605002}}].

\bibitem{MiniBooNEDM:2018cxm}
{\scshape MiniBooNE DM} collaboration, A.~A. Aguilar-Arevalo et~al.,
  \emph{{Dark Matter Search in Nucleon, Pion, and Electron Channels from a
  Proton Beam Dump with MiniBooNE}},
  \href{http://dx.doi.org/10.1103/PhysRevD.98.112004}{\emph{Phys. Rev. D} {\bf
  98} (2018) 112004}, [\href{http://arxiv.org/abs/1807.06137}{{\tt
  1807.06137}}].

\bibitem{CHARM-II:1989nic}
{\scshape CHARM-II} collaboration, K.~De~Winter et~al., \emph{{A Detector for
  the Study of Neutrino - Electron Scattering}},
  \href{http://dx.doi.org/10.1016/0168-9002(89)91190-X}{\emph{Nucl. Instrum.
  Meth. A} {\bf 278} (1989) 670}.

\bibitem{CHARM-II:1994dzw}
{\scshape CHARM-II} collaboration, P.~Vilain et~al., \emph{{Precision
  measurement of electroweak parameters from the scattering of muon-neutrinos
  on electrons}},
  \href{http://dx.doi.org/10.1016/0370-2693(94)91421-4}{\emph{Phys. Lett. B}
  {\bf 335} (1994) 246--252}.

\bibitem{Ball:1980ojt}
R.~Ball et~al., \emph{{The Neutrino Beam Dump Experiment at Fermilab (E613)}},
  {\emph{eConf} {\bf C801002} (1980) 172--174}.

\bibitem{SHiP:2015vad}
{\scshape SHiP} collaboration, M.~Anelli et~al., \emph{{A facility to Search
  for Hidden Particles (SHiP) at the CERN SPS}},
  \href{http://arxiv.org/abs/1504.04956}{{\tt 1504.04956}}.

\bibitem{DUNE:2015lol}
{\scshape DUNE} collaboration, R.~Acciarri et~al., \emph{{Long-Baseline
  Neutrino Facility (LBNF) and Deep Underground Neutrino Experiment (DUNE)}:
  {Conceptual Design Report, Volume 2: The Physics Program for DUNE at LBNF}},
  \href{http://arxiv.org/abs/1512.06148}{{\tt 1512.06148}}.

\bibitem{DUNE:2017pqt}
{\scshape DUNE} collaboration, B.~Abi et~al., \emph{{The Single-Phase ProtoDUNE
  Technical Design Report}},  \href{http://arxiv.org/abs/1706.07081}{{\tt
  1706.07081}}.

\bibitem{Fortin:2011hv}
J.-F. Fortin and T.~M.~P. Tait, \emph{{Collider Constraints on
  Dipole-Interacting Dark Matter}},
  \href{http://dx.doi.org/10.1103/PhysRevD.85.063506}{\emph{Phys. Rev. D} {\bf
  85} (2012) 063506}, [\href{http://arxiv.org/abs/1103.3289}{{\tt 1103.3289}}].

\bibitem{Zhang:2022ijx}
Y.~Zhang, M.~Song and L.~Chen, \emph{{Dark states with electromagnetic form
  factors at electron colliders}},
  \href{http://dx.doi.org/10.1103/PhysRevD.107.055023}{\emph{Phys. Rev. D} {\bf
  107} (2023) 055023}, [\href{http://arxiv.org/abs/2208.08142}{{\tt
  2208.08142}}].

\bibitem{Kling:2022ykt}
F.~Kling, J.-L. Kuo, S.~Trojanowski and Y.-D. Tsai, \emph{{FLArE up dark
  sectors with EM form factors at the LHC forward physics facility}},
  \href{http://dx.doi.org/10.1016/j.nuclphysb.2023.116103}{\emph{Nucl. Phys. B}
  {\bf 987} (2023) 116103}, [\href{http://arxiv.org/abs/2205.09137}{{\tt
  2205.09137}}].

\bibitem{Anchordoqui:2021ghd}
L.~A. Anchordoqui et~al., \emph{{The Forward Physics Facility: Sites,
  experiments, and physics potential}},
  \href{http://dx.doi.org/10.1016/j.physrep.2022.04.004}{\emph{Phys. Rept.}
  {\bf 968} (2022) 1--50}, [\href{http://arxiv.org/abs/2109.10905}{{\tt
  2109.10905}}].

\bibitem{Chu:2019rok}
X.~Chu, J.-L. Kuo, J.~Pradler and L.~Semmelrock, \emph{{Stellar probes of dark
  sector-photon interactions}},
  \href{http://dx.doi.org/10.1103/PhysRevD.100.083002}{\emph{Phys. Rev. D} {\bf
  100} (2019) 083002}, [\href{http://arxiv.org/abs/1908.00553}{{\tt
  1908.00553}}].

\bibitem{Chang:2019xva}
J.~H. Chang, R.~Essig and A.~Reinert, \emph{{Light(ly)-coupled Dark Matter in
  the keV Range: Freeze-In and Constraints}},
  \href{http://dx.doi.org/10.1007/JHEP03(2021)141}{\emph{JHEP} {\bf 03} (2021)
  141}, [\href{http://arxiv.org/abs/1911.03389}{{\tt 1911.03389}}].

\bibitem{ILCInternationalDevelopmentTeam:2022izu}
{\scshape ILC International Development Team} collaboration, A.~Aryshev et~al.,
  \emph{{The International Linear Collider: Report to Snowmass 2021}},
  \href{http://arxiv.org/abs/2203.07622}{{\tt 2203.07622}}.

\bibitem{Robson:2018zje}
A.~Robson and P.~Roloff, \emph{{Updated CLIC luminosity staging baseline and
  Higgs coupling prospects}},  \href{http://arxiv.org/abs/1812.01644}{{\tt
  1812.01644}}.

\bibitem{Adli:2025swq}
E.~Adli et~al., \emph{{The Compact Linear e$^+$e$^-$ Collider (CLIC)}},
  \href{http://arxiv.org/abs/2503.24168}{{\tt 2503.24168}}.

\bibitem{Greene:1986bm}
B.~R. Greene, K.~H. Kirklin, P.~J. Miron and G.~G. Ross, \emph{{A Three
  Generation Superstring Model. 1. Compactification and Discrete Symmetries}},
  \href{http://dx.doi.org/10.1016/0550-3213(86)90057-X}{\emph{Nucl. Phys. B}
  {\bf 278} (1986) 667--693}.

\bibitem{Craig:2019wmo}
N.~Craig, M.~Jiang, Y.-Y. Li and D.~Sutherland, \emph{{Loops and Trees in
  Generic EFTs}}, \href{http://dx.doi.org/10.1007/JHEP08(2020)086}{\emph{JHEP}
  {\bf 08} (2020) 086}, [\href{http://arxiv.org/abs/2001.00017}{{\tt
  2001.00017}}].

\bibitem{Nicrosini:1989pn}
O.~Nicrosini and L.~Trentadue, \emph{{Transverse Degrees of Freedom in {QED}
  Evolution}},
  \href{http://dx.doi.org/10.1016/0370-2693(89)90699-0}{\emph{Phys. Lett. B}
  {\bf 231} (1989) 487--491}.

\bibitem{Montagna:1995wp}
G.~Montagna, O.~Nicrosini, F.~Piccinini and L.~Trentadue, \emph{{Invisible
  events with radiative photons at LEP}},
  \href{http://dx.doi.org/10.1016/0550-3213(95)00399-D}{\emph{Nucl. Phys. B}
  {\bf 452} (1995) 161--172}, [\href{http://arxiv.org/abs/hep-ph/9506258}{{\tt
  hep-ph/9506258}}].

\bibitem{Habermehl:2020njb}
M.~Habermehl, M.~Berggren and J.~List, \emph{{WIMP Dark Matter at the
  International Linear Collider}},
  \href{http://dx.doi.org/10.1103/PhysRevD.101.075053}{\emph{Phys. Rev. D} {\bf
  101} (2020) 075053}, [\href{http://arxiv.org/abs/2001.03011}{{\tt
  2001.03011}}].

\bibitem{Berends:1983fs}
F.~A. Berends and R.~Kleiss, \emph{{Distributions in the Process e+ e-
  ---{\ensuremath{>}} e+ e- (Gamma)}},
  \href{http://dx.doi.org/10.1016/0550-3213(83)90558-8}{\emph{Nucl. Phys. B}
  {\bf 228} (1983) 537--551}.

\bibitem{Tobimatsu:1985vd}
K.~Tobimatsu and Y.~Shimizu, \emph{{Radiative Correction to $e^+ e^- \to e^+
  e^-$ in the Electroweak Theory. 1. Cross-sections for Hard Photon Emission}},
  \href{http://dx.doi.org/10.1143/PTP.74.567}{\emph{Prog. Theor. Phys.} {\bf
  74} (1985) 567}.

\bibitem{Habermehl:2018yul}
M.~Habermehl, \emph{{Dark Matter at the International Linear Collider}}.
\newblock PhD thesis, Hamburg U., Hamburg, 2018.
\newblock 10.3204/PUBDB-2018-05723.

\bibitem{Alwall:2014hca}
J.~Alwall, R.~Frederix, S.~Frixione, V.~Hirschi, F.~Maltoni, O.~Mattelaer
  et~al., \emph{{The automated computation of tree-level and next-to-leading
  order differential cross sections, and their matching to parton shower
  simulations}}, \href{http://dx.doi.org/10.1007/JHEP07(2014)079}{\emph{JHEP}
  {\bf 07} (2014) 079}, [\href{http://arxiv.org/abs/1405.0301}{{\tt
  1405.0301}}].

\bibitem{Blaising:2021vhh}
{\scshape CLICdp} collaboration, J.-J. Blaising, P.~Roloff, A.~Sailer and
  U.~Schnoor, \emph{{Physics performance for Dark Matter searches at
  $\sqrt{s}=$ 3 TeV at CLIC using mono-photons and polarised beams}},
  \href{http://arxiv.org/abs/2103.06006}{{\tt 2103.06006}}.

\bibitem{Barducci:2025twe}
D.~Barducci, D.~Buttazzo, A.~Dondarini, R.~Franceschini, G.~Marino, F.~Mescia
  et~al., \emph{{Scalar Rayleigh Dark Matter: current bounds and future
  prospects}}, \href{http://dx.doi.org/10.1007/JHEP06(2025)171}{\emph{JHEP}
  {\bf 06} (2025) 171}, [\href{http://arxiv.org/abs/2501.09073}{{\tt
  2501.09073}}].

\end{thebibliography}\endgroup
\end{document}